\documentclass{aa}

\pdfoutput=1
\usepackage{txfonts}
\usepackage[sort&compress]{natbib}
\bibpunct{(}{)}{;}{a}{}{,} 
\usepackage{amssymb}
\usepackage{graphicx}
\usepackage{url}
\usepackage{mathrsfs}
\usepackage{color}

\begin{document}

\title{Type Ia Supernova host galaxies as seen with IFU spectroscopy\thanks{Based on observations collected at the Centro Astron\'omico Hispano Alem\'an (CAHA) at Calar Alto, operated jointly by the Max-Planck Institut f\"ur Astronomie and the Instituto de Astrof\'isica de Andaluc\'ia (CSIC)}}
\author{Stanishev, V.\inst{\ref{inst1}}
\and Rodrigues, M.\inst{\ref{inst2},\ref{inst3}}
\and Mour\~ao, A.\inst{\ref{inst1}} 
\and Flores, H.\inst{\ref{inst2}}
 }

\institute{
CENTRA - Centro Multidisciplinar de Astrof\'isica, Instituto Superior T\'ecnico, Av.
Rovisco Pais 1, 1049-001 Lisbon, Portugal\label{inst1}
\and GEPI , Observatoire de Paris, CNRS, University Paris Diderot, 5 Place Jules Janssen, 92195 Meudon, France \label{inst2}
\and European Southern Observatory, Alonso de Cordova 3107, Casilla 19001, Vitacura, Santiago, Chile\label{inst3}}

\date{Received data Accepted date}

\abstract{
Type Ia Supernovae (SNe~Ia) have been widely used in cosmology as distance indicators. 
However, to fully exploit their potential in cosmology, a better control over systematic uncertainties 
is required. Some of the uncertainties are related to the unknown nature of the SN~Ia progenitors. 
}
{
We aim to test the use of integral field unit (IFU) spectroscopy for correlating the properties of  
nearby SNe~Ia  with the properties of their host galaxies at the location of the SNe.
The results are to be compared with those obtained from an analysis of the total host spectrum.
The goal is to explore this path of constraining the nature of the SN~Ia progenitors and further 
improve the use of SNe~Ia in cosmology.
}
{
We used the wide-field IFU spectrograph PMAS/PPAK at the 3.5m telescope of Calar Alto Observatory
 to observe six nearby spiral galaxies that hosted SNe~Ia. Spatially resolved 2D maps 
 of the properties of the ionized gas and the stellar populations were derived. 
}	
{
Five of the observed galaxies have an ongoing star formation rate of 1-5 $M_\odot$\,yr$^{-1}$ and mean stellar population 
ages $\sim5$ Gyr. The sixth galaxy shows no star formation and has an about 12 Gyr old stellar population. 
All galaxies have stellar masses larger than $2\times10^{10}M_\odot$ and metallicities above solar.
Four galaxies show negative radial metallicity gradients of the ionized gas up to $-0.058$ dex\,kpc$^{-1}$ and one has 
nearly uniform metallicity with a possible shallow positive slope. The stellar components show shallower negative metallicity gradients up to $-0.03$ dex\,kpc$^{-1}$.
 We find no clear correlation between the properties of the galaxy and those of the supernovae, 
which may be because of  the small ranges spanned by the galaxy parameters. 
 However, we note that the Hubble residuals are on average positive while negative Hubble 
 residuals are expected for SNe~Ia in massive hosts such as the galaxies in our sample. 
}
{
The IFU spectroscopy on 4-m telescopes is a viable technique for studying host galaxies of  
nearby SNe~Ia. It allows one to correlate the supernova properties with the properties
of their host galaxies at the projected positions of the supernovae.  Our current sample of six galaxies is too small
 to draw conclusions about  the SN~Ia progenitors or correlations with the galaxy properties, 
 but the ongoing CALIFA IFU survey  will provide a solid basis to 
   exploit this technique more and improve our understanding of SNe~Ia as cosmological standard candles.
}

\keywords{Galaxies: general -- Galaxies: abundances -- (Stars:) supernovae: general}

\maketitle
\titlerunning{3D spectroscopy of SNe~Ia host galaxies}
\authorrunning{Stanishev et al.}

\section{Introduction}

More than a decade ago the observations of Type Ia Supernovae (SNe~Ia) 
led to the discovery of the accelerating expansion of the Universe 
and the need for an unknown repulsive force to drive it 
\citep{1998AJ....116.1009R,p99}. Understanding the nature of this force 
-- now dubbed "dark energy" -- is an outstanding goal of
astrophysics and cosmology. It is now well-understood that no single observational technique will be able to 
achieve this alone \citep[e.g.][]{detf}.
The combined constraints of many techniques will be needed to measure the equation-of-state parameter 
of the dark energy (the ratio between the pressure 
and the density) and eventually its evolution over the cosmic time. 
At present, SNe~Ia is the most matured and well-understood technique to accurately trace the 
cosmic expansion history and will continue to play an essential role in future cosmological
experiments \citep{detf}. However, the SNe~Ia technique is affected by several
systematic uncertainties, which need to be reduced to a level below $\sim2$\%
to differentiate between different dark energy models. 
  
 The use of SNe~Ia in Cosmology relies   
on the \emph{empirically} established tight relation
between their light curve width and peak luminosity, which allows one to
measure the luminosity distance with an accuracy of $\sim7$\% \citep{phil99},
and on the \emph{assumption} that the (standardized) peak luminosity of SNe Ia 
does not change over the cosmic time.
There is now observational evidence that the slope of the "light curve shape - peak luminosity" relation 
does not depend on the redshift or the host galaxy mass \citep{2011ApJS..192....1C,2011ApJ...737..102S}. 
However, \citet{2010MNRAS.406..782S} and \citet{2010ApJ...715..743K} have found that
the offsets of the SN~Ia peak magnitudes from the best-fitting Hubble line 
(from now on Hubble residuals or HR) correlate with the host stellar mass. Together with the 
mass-metallicity relation for galaxies \citep[e.g.,][]{2004ApJ...613..898T} and the overall increase of the metal content of the universe with the cosmic time, 
this may be an indication for possible luminosity evolution of SNe Ia at a level of 0.05-0.10 mag.

It is now generally
agreed that SNe~Ia are the result of thermonuclear disruption of
carbon/oxygen (C/O) white dwarfs (WD),  which ignite explosively when they
approach the Chandrasekhar limit $M_\mathrm{Ch}\sim1.38M_{\sun}$ \citep{1960ApJ...132..565H}.
However, there has been little observational evidence of the exact
evolutionary scenario that leads to the explosion. 
C/O WDs are the end product of the evolution of 
stars with masses $\sim1.5-7M_{\sun}$ \citep[e.g., see][]{1980ApJ...237..111B,1999ApJ...524..226D}.
The upper mass limit for C/O WDs is  $\sim1.1M_{\sun}$ 
\citep[e.g., see][]{1987A&A...188...74W,1999ApJ...524..226D,2009ApJ...692.1013S} and therefore
 a mechanism that allows the WD to gain additional mass of at least 
$\sim0.3-0.4M_{\sun}$ is needed.
In the single degenerate (SD) scenario the WD accretes mass from a
non-degenerate companion star in a binary system \citep{1973ApJ...186.1007W}. However, the exact physical mechanism 
of the WD mass growth has not yet been identified.
In the double-degenerate (DD) scenario two C/O WDs in a binary merge after loosing orbital angular momentum by
 gravitational wave radiation \citep{1984ApJ...284..719I,1984ApJ...277..355W,1985ASSL..113....1P}. 
While considered the most viable, both scenarios have considerable
uncertainties \citep[see, e.g.][]{hille00,2011arXiv1111.4492M}.

 The numerical simulations of thermonuclear SN~Ia explosions suggest 
that the properties of the exploding WD may significantly influence the peak luminosity, the
"light curve width - luminosity" relation and colors of the resulting supernovae
 \citep[e.g.,][]{hof98,1999ApJ...522L..43U,2001ApJ...557..279D,2006A&A...453..203R,2009Natur.460..869K,2010ApJ...711L..66B}. 
On the other hand,  the properties of the WD just before the ignition  
(the central density, metallicity and C/O ratio)
 are sensitive to the properties and the evolution of its progenitor binary star, and to the
 subsequent WD mass growth mechanism.
For example, the SD channel can produce $M_\mathrm{Ch}$ WDs with slightly different structure and chemical
composition depending on the mass of the WD at the moment when the accretion started \citep[e.g.,][]{2001ApJ...557..279D}.
In the DD scenario, the outcome of the merger may depend on the mass ratio of the two WDs.
 In addition, one may expect that some properties
of  SN~Ia progenitor stars will evolve with cosmic time, e.g. metallicity. 
 Therefore,
possible evolution of the properties of SN~Ia progenitors  or,  if more than one evolutionary channel exist, evolution 
of their relative contribution to the SNe~Ia population, may introduce systematic
uncertainties in SN~Ia cosmology and potentially bias the cosmological results
from the future large SN surveys \citep[e.g., see][]{2008ApJ...684L..13S,2008JCAP...02..008N}. 

To date no progenitor of a SN~Ia has been unambiguously identified and/or observed and
information about the SNe~Ia progenitors has been inferred indirectly.
\cite{2005A&A...433..807M}  and \cite{2005ApJ...629L..85S} studied the SN~Ia rate as 
a function of redshift and host galaxies properties. Both studies found that the SN~Ia rate
depends on both the on-going star formation rate (SFR) and total galaxy stellar mass. 
This result was also confirmed by others 
\citep{2006ApJ...648..868S,2006AJ....132.1126N,2006MNRAS.370..773M,pritchet08s,dahlen08,2011MNRAS.412.1508M} and 
appears to suggest that at least part of SNe~Ia are associated with the young stellar population
capable of producing SN~Ia with short delay time $\leq400$ Myr.
\cite{2008PASJ...60.1327T} and \cite{2010ApJ...722.1879M} have shown that the 
delay times from star formation to SN~Ia explosions between the shortest time probed $<400$ Myr and 10 Gyr 
are distributed as a power law with slope $\sim-1$. This delay time distribution (DTD) strongly favors the DD scenario. 
The SD scenario may also explain this DTD \citep{2008ApJ...683L.127H} but the efficiency of the 
symbiotic channel (WD+red giant) needs to be significantly increased \citep[see, e.g.,][]{2011arXiv1111.4492M}.
The early discovery of \object{SN~2011fe} in \object{M101}, the nearest SN Ia in 25 years, provided
the first real possibility to constrain the properties of a progenitor star of an SN Ia
\citep{2012ApJ...744L..17B,2012ApJ...750..164C,2011Natur.480..344N}. The results reinforce the conclusion that 
the exploding star is a C/O WD and seem to rule out all but a degenerate star as its companion, thus favoring
the DD scenario. On the other hand, based on high-resolution spectroscopy of a sample of nearby SNe Ia \cite{2011Sci...333..856S}
favor the SD scenario.

Many studies have shown that the intrinsically luminous SNe tend to 
occur in star-forming hosts, while the faint SNe prefer passive ones
\citep[e.g.,][]{1996AJ....112.2391H,2000AJ....120.1479H,
2005ApJ...634..210G,2008ApJ...685..752G,2009ApJ...691..661H,
2009ApJ...707.1449N,2006ApJ...648..868S,2010MNRAS.406..782S,2010ApJ...715..743K,2010AJ....140..804B,2009ApJ...707...74R}.  
\citet{2008ApJ...685..752G} found that the Hubble residuals correlate with the global host metallicity. 
 \citet{2010MNRAS.406..782S} and \citet{2010ApJ...715..743K} found such a correlation with the host stellar mass. However, 
\citet{2009ApJ...691..661H},  who used the galaxy stellar mass as a proxy for the metallicity, found no such correlation and suggested that instead the progenitor age may be a more important parameter.

 All studies of SN Ia hosts galaxies conducted so far, except that by 
\citet{2009ApJ...707...74R}, were based on an analysis of the global photometric or spectroscopic properties of the 
host galaxies.
In this paper we take a different approach. We use for the first time integral field unit (IFU) spectroscopy 
at intermediate spectral resolution to study a sample of host galaxies of local SNe~Ia ($z\sim0.02$). This 
 approach has an advantage over the previous studies because it allows us to derive spatially-resolved 
 two-dimensional (2D) maps of host galaxy properties, e.g.  the heavy element abundance in the
interstellar medium (ISM). The intermediate spectral resolution 
makes it also possible to use full-spectrum fitting techniques to derive 2D maps of the properties of the stellar populations.
The main objective of this pilot work is to test the methodology to correlate the properties of the SNe with  the properties of the
gas and the stellar populations \emph{at the location of the SN explosion}, in addition to the global host properties.
By analyzing the properties of the stellar populations  we also aim to constrain the nature of the SNe~Ia progenitors. 

Throughout the paper we assume the concordance cosmological
model with $\Omega_M=0.27$, $\Omega_\Lambda=0.73$, $w=-1$ and $h=0.71$.


\begin{table*}[!t]

\caption{Supernovae and details of their host galaxies: morphological type, Milky Way dust reddening,
offsets from the host nucleus, de-projected galactocentric distance, inclination, and position angle.} 
\label{t:hosts}
\begin{tabular}{@{}lllccccccc@{}}
\hline
\hline\noalign{\smallskip}
SN   	&    Host      	& Type\tablefootmark{a}   & z\tablefootmark{b}        & $E(B-V)_{MW}$\tablefootmark{c} &   	RA	offset	& DEC offset & DGD\tablefootmark{b} & $i$\tablefootmark{b} & PA\tablefootmark{b} \\
	    &      	    	&      		 &          &  &             [arcsec]    & [arcsec]& [kpc]& [deg] & [deg] \\
\hline\noalign{\smallskip}
1999dq 	&  NGC 976    	& SAb       &  0.0144	& 0.110 & $-$4.0 	& $-$6.0  & 2.4 & 36.5 & 77.6 \\ 
1999ej 	&  NGC 495	    & SB(s)0/a  &  0.0137 & 0.072  	&  +18.0	& $-$20.0 & 7.7 & 44.0 & 47.6 \\ 
2001fe 	&  UGC 5129     & SBbc      &  0.0134 & 0.022  	& $-$13.5	& $-$0.1  & 3.9 & 46.0 & 14.1	\\ 
2006te 	&  CGCG 207-042 & SB(r)bc   &  0.0315	& 0.046 & $-$5.5    & $-$1.7  & 4.9 & 39.4 & 69.9	\\ 
2007A  	&  NGC 105 NED02& SBc	    &  0.0173 & 0.073  	& $-$1.2	& +10.1   & 3.6 & 36.4 & 77.3	\\ 
1997cw 	&  NGC 105 NED02& SBc	    &  0.0173 & 0.073  	&   +8.0    & +4.0 	  & 4.0 & 36.4 & 77.3 \\ 
2007R  	&  UGC 4008 NED01 & SAa     &  0.0308 & 0.047  	& $-$1.9 	& $-$3.9  & 3.4 & 48.1 & 76.4  \\ 
\hline 
\end{tabular}\\
\tablefoottext{a}{based on SDSS pseudo-color images.}
\tablefoottext{b}{this work. Derived from analysis of the H$\alpha$ velocity field, except for 
the host of \object{SN 1999ej}, for which the stellar velocity field was used.}
\tablefoottext{c}{from \cite{ebv}.}

\end{table*}

\section{Observations and data reduction}

\subsection{Target selection}

The list of targets for this program was selected from a sample of spiral  galaxies that hosted 
 SNe~Ia for which the important parameters such as luminosity, 
extinction, intrinsic color indices, luminosity decline rate, and deviation from 
the Hubble diagram have been accurately measured.
The galaxies were carefully examined and selected to fulfill four additional criteria: 

\begin{enumerate}

\item to have angular size $\simeq$40-60 arcsec; 

\item to be nearly face-on; 

\item the SN lies on a high surface brightness location in the galaxy;

\item be observable at airmass less that 1.3 to minimize the effect of the differential atmospheric
refraction.

\end{enumerate}

The first two requirements maximize the use of the large field-of-view
(FOV) of the IFU instrument and minimize the projection effects when correlating the SN 
and its local host galaxy properties. The third requirement ensures that we 
will obtain a good signal-to-noise ratio (S/N) of the spectra at the location of the SN, and in particular 
will allow us to access the absorption and emission line spectra. From the 
large sample of galaxies six objects were observed. The galaxy details and the SNe offset from the nucleus 
 are given in Table\,\ref{t:hosts}. All SNe are normal SNe~Ia, except for 
\object{SN~1999dq}, which was classified as a peculiar 1991T-like event \citep{1999IAUC.7250....1J}. 
 Table\,\ref{t:snprop} gives the  SALT2  $x_1$ and $C$ parameters of the supernova light curves \citep{2007A&A...466...11G} taken from \cite{2010ApJ...716..712A}, the offset from the best-fit Hubble line $\Delta\mu$, and the $\Delta M_{15}$ parameter, which shows how much the SN $B$-band magnitude has declined during the first 15 days after the time of the $B$ band maximum. 
 $x_1$ and $C$ are parameters related to the SN
light curve shape and $B-V$ color index at maximum, respectively. They are used to standardize the 
observed  $B$-band peak magnitude $B_{\rm obs}$ via the relation 
$B_{\rm std}=B_{\rm obs}+\alpha x_1 - \beta C$, with $\alpha=0.121$ and 
$\beta=2.51$ as per \cite{2010ApJ...716..712A}. $\Delta\mu$ was computed after first correcting the 
redshifts of the SNe for  large-scale coherent galaxy motions in the local universe based on the 
models of \cite{2004MNRAS.352...61H}. The accuracy of this correction is estimated to be $\sim150$\,km\,s$^{-1}$ and 
  a random peculiar velocity of 150\,km\,s$^{-1}$ is added to the uncertainty of  $\Delta\mu$.

It should be noted that for this pilot project the selection criteria are solely optimized to maximize the 
quality of the observations and facilitate the analysis. 
We focus on late-type galaxies because one of our goals is 
 to correlate the SN properties with the properties of the ISM determined from the ionized gas.
This leads to strong biases, however, e.g. the galaxies in our
sample are bright, massive, and likely metal-rich.

\subsection{Observations}

The six galaxies  were observed 
on November 14 and 15, 2009 at the 3.5m
telescope of the Calar Alto observatory using the Potsdam Multi-Aperture
Spectrograph \citep[PMAS,][]{2005PASP..117..620R} in the PPAK mode
\citep{2004AN....325..151V,2006PASP..118..129K}. 
The atmospheric conditions were variable with occasional thin clouds interrupting the observations. 
The seeing varied between 1.5\arcsec and 2.2\arcsec. 

The PMAS instrument is equipped with a 
4k$\times$4k E2V\#231 CCD. We used a set-up with the 600 lines\,mm$^{-1}$ grating 
V600 and 2$\times$2 binned CCD, which
provided a wavelength range of $\sim$3700-7000\AA\ with a spectral resolution of
$\sim5.5$\AA. 
The PPAK fiber bundle of PMAS consists of 382  
fibers with 2.7\arcsec diameter each, 331 of which (science fibers) are ordered
in a single hexagonal bundle that covers a FOV of 
72\arcsec$\times$64\arcsec. 
Thirty-six additional fibers
 form six mini-bundles (sky-bundles), which are evenly
distributed along a circle of $\sim90$\arcsec radius and face the edges of the
central hexagon  \citep[see Fig.5 in][]{2006PASP..118..129K}. 
The remaining 15 fibers are used for calibration and can only
be illuminated with the PMAS internal calibration unit.
For a detailed description of the PPAK fiber
bundle we refer the reader to \cite{2006PASP..118..129K}. Some details that are relevant for the
data reduction are also given in Appendix~\ref{ap:instr}.

For each object three 1800-sec long exposures were obtained. 
Because the filling factor of a single PPAK exposure is $\sim$65\%, we adopted 
a dithering pattern with the 
second and the third exposures offset by $\Delta$(R.A.,Decl.)=(1.56, 0.78)
and (1.56, $-$0.78) arcsec with respect to the first exposure to 
ensure that every point within the FOV was spectroscopically sampled. 
Before and after the science exposures,
spectra of HgNe and continuum halogen lamp were obtained to wavelength-calibrate and trace the spectra. 
 The spectrophotometric standard stars 
\object{Feige\,34} and \object{BD+25\,3941} were observed 
 to measure the sensitivity function of the instrument.
In addition, series of exposures of blank sky regions were obtained
during twilight and were used to equalize the fiber-to-fiber throughput
variations.

\subsection{Data reduction}

The pre-reduction of the CCD images  was performed with IRAF\footnote{IRAF is
distributed by the National Optical Astronomy Observatories, which are operated
by the Association of Universities for Research in Astronomy, Inc., under
cooperative agreement with the National Science Foundation.} and the rest of the
reduction  with our own programs written in IDL. Each individual science pointing 
was reduced independently. After the standard CCD reduction steps of bias subtraction,
flat-field correction and removal of cosmic ray hits, the spectra were traced, extracted, wavelength- and
flux-calibrated, and finally sky-subtracted. At the final step the three pointings were 
combined into a final 3D data-cube, taking into account the differential atmospheric diffraction.
The full details of the data reduction are given in Appendix~\ref{ap:reduction}.

\begin{table}[!t]
\caption{SALT2 \citep{2007A&A...466...11G} $x_1$ and $C$ parameters of the SNe from \cite{2010ApJ...716..712A},
the offset from the best-fit Hubble line $\Delta\mu$, and the $\Delta M_{15}$ parameter. The uncertainties of the 
parameters are given in the parentheses.} 
\label{t:snprop}
\begin{tabular}{@{}lcccc@{}}
\hline
\hline\noalign{\smallskip}
 SN    &  $x_1$       &  C           & $\Delta\mu$  & $\Delta M_{15}$\tablefootmark{a} \\
\hline\noalign{\smallskip}
1999dq &    0.89 (0.12) &    0.13 (0.01) & $-$0.33 (0.09) &   0.96  \\
1999ej & $-$2.08 (0.43) &    0.07 (0.05) &    0.46 (0.20) &   1.49  \\
2001fe &    0.41 (0.18) &    0.03 (0.02) & $-$0.06 (0.09) &   1.03  \\
2006te & $-$0.36 (0.18) & $-$0.04 (0.02) &    0.16 (0.08) &   1.15  \\
1997cw\tablefootmark{b} &    0.79 (0.25) &    0.40 (0.03) &   $-$0.08 (0.13) &   0.97  \\
2007A  & $-$0.04 (0.14) &    0.18 (0.01) &    0.15 (0.07) &   1.10  \\
2007R  & $-$1.76 (0.16) & $-$0.07 (0.02) &    0.23 (0.07) &   1.42   \\
\hline 
\end{tabular}\\
\tablefoottext{a}{calculated from $x_1$ with the relation given in \cite{2007A&A...466...11G};}
\tablefoottext{b}{the first photometric observation was taken  $\sim$15 days past maximum and the
photometric parameters are rather uncertain. This SN was included in the analysis because it is in
the same host as \object{SN~2007A}.}

\end{table}

\begin{figure}[!t]
\includegraphics [width=8.8cm]{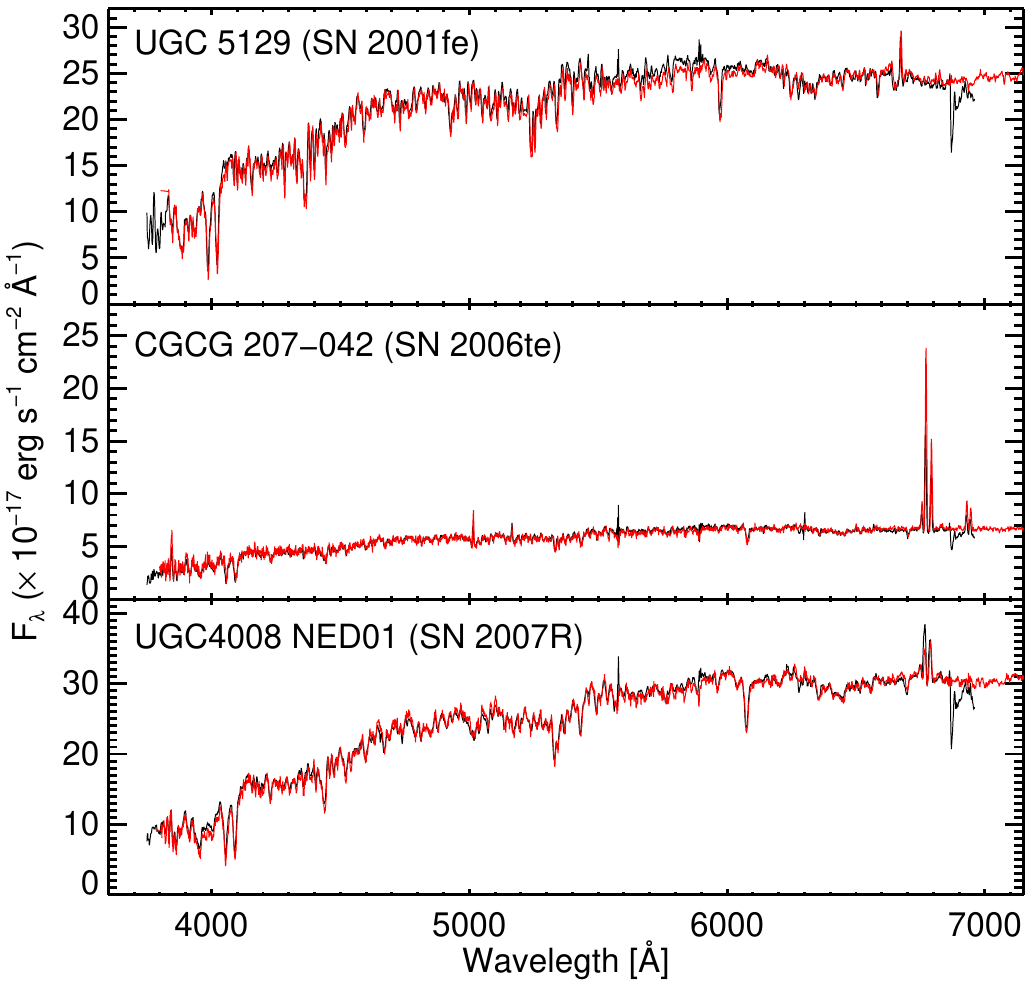} 
\caption{Comparison between SDSS spectra (red) and spectra extracted from our data-cubes within an aperture
of 3\arcsec diameter centered on the galaxy nuclei (black).}
\label{f:sdss}
\end{figure}

Three of the galaxies in our sample also have SDSS spectra. This allowed us 
to check the \emph{relative} flux calibration of our spectroscopy. Spectra within an 
aperture of 3\arcsec\ diameter centered on the galaxy nucleus were extracted from the data-cubes to
emulate the SDSS spectra. The comparison, after our spectra were scaled to match the flux level 
of the SDSS spectra, is shown in Fig.\,\ref{f:sdss}. It demonstrates  that the \emph{relative}
 flux calibration of our spectra is excellent and matches SDSS to within a few percent.

The absolute flux scale of the data-cubes was set using the SDSS 
imaging. SDSS $g$ and $r$ magnitudes of the galaxies were 
 computed within an aperture of 20\arcsec\ diameter. Spectra within the same aperture size 
 were extracted from the data-cubes and synthetic $g$ and $r$ magnitudes were computed. 
 The $g$ and $r$ scale factors that provided the match of the synthetic magnitudes to the observed 
 ones were computed and the average of the two was applied to the data-cubes. 
 We note that the  $g$ and $r$ scale factors coincided to within 3\%, 
which additionally supports our conclusion that the \emph{relative} flux calibration is accurate.

\section{Data analysis}
\label{sec:analysis}

 The individual spectra in the data-cubes were analyzed to derive 2D maps of the properties of the
galaxies. This included the  properties of the ionized gas and the stellar populations. The properties of the 
galaxies at the SN position and the galaxy center  were derived by
interpolating the 2D maps. The galaxy centers were computed from the data-cubes and the SN positions were
computed with respect to it, using the offsets quoted in the discovery IAU circulars and \cite{jha44}.

As previously mentioned, one of the main goals of this study is to test the feasibility of
using IFU spectroscopy to compare the properties of the host as derived from integrated
spectroscopy to those derived from spatially resolved spectroscopy. For this purpose, we also analyzed for each galaxy  
the total spectrum formed by simply summing all spaxel spectra in the data-cube.
This simulates an observation of the same galaxy with long-slit spectroscopy as is performed for high-redshift galaxies.

For each galaxy the analysis was also performed on azimutally averaged spectra at several de-projected galactocentric radii.
 This was performed as an alternative way to derive the radial dependence of the galaxy properties such as the metallicity. 
 To compute the azimutally averaged spectra we first computed the de-projected galactocentric distance of each spaxel with the position angle and inclination (Table~\ref{t:hosts})
computed from the analysis of the H$\alpha$ velocity maps\footnote{For \object{NGC~495} the stellar 
velocity map was used because this galaxy shows no emission lines.} (see Sec.~\ref{sec:gasvel}). The spectra
were corrected to rest-frame wavelength with the stellar velocities estimated from the fits
to the absorption line spectrum (Sec.~\ref{sec:starfits}). Finally, for each galaxy the spectra
within several (4 to 6) radial bins were averaged. Because the stellar velocities were used to 
correct the spectra to rest frame, we used only the spectra that had a sufficiently high S/N to allow fitting with {\tt STARLIGHT}.

All quoted uncertainties of the derived quantities are statistical and do not include systematic and 
intrinsic uncertainties of the methods, which will be additionally discussed when appropriate. 
In the next sections we present the main steps in the data analysis.

\subsection{Properties of the ionized gas}

The presence of nebular emission lines in the galaxy spectra allows us to study the properties of the ionized gas
such as its oxygen abundance and ionization state, and to derive other important properties such as the
star formation rate, dust extinction, etc.
Some of the methods used to derive these quantities, for example  the strong line methods to estimate the gas metallicity, 
can only be applied if the
ionization source is exclusively arising from the stellar radiation.  For this reason and  to search for 
possible AGN contamination, we used the diagnostic
diagram [\ion{O}{iii}]\,$\lambda$5007/H$\beta$ \emph{vs.} [\ion{N}{ii}]\,$\lambda$6584/H$\alpha$ \citep[BPT diagram;][]{1981PASP...93....5B} (Fig.\,\ref{f:bpt}).
The central spaxels that fall in the AGN area of the diagnostic diagram according to the  
\cite{2003MNRAS.346.1055K} criterion were excluded from the relevant parts of the analysis.

\subsubsection{Emission line fluxes}

 Five of the six galaxies in our sample show strong
nebular emission lines.
The fluxes of the prominent emission lines
[\ion{O}{ii}]\,$\lambda$3727, H$\beta$,
[\ion{O}{iii}]\,$\lambda\lambda$4959/5007,  H$\alpha$, and
[\ion{N}{ii}]\,$\lambda\lambda$6549/6584 were used in the analysis.
Whenever possible, the [\ion{S}{ii}]\,$\lambda\lambda$6716/6731 lines
were also measured. 
In the spectra of galaxies the emission lines are superimposed on the underlying stellar
absorption spectrum. The stellar absorption lines can bias the measurement of the
emission line fluxes, an effect that is especially prominent in the H$\beta$ line (Fig.~\ref{f:contsub}). Therefore,
to measure the emission line fluxes accurately, the stellar absorption spectrum needs to be
subtracted first. For this we used the STARLIGHT software
\citep{2005MNRAS.358..363C}. 
All spectra that had S/N greater than 5 at
$\sim4600$\AA\ were fitted with STARLIGHT and the emission line fluxes were measured on the
continuum-subtracted spectrum. For the remaining the spectra the measurements were
made without continuum subtraction. Each emission line was fitted with a
single Gaussian plus a linear term, and the area under the Gaussian was taken as flux estimate.
Details of the adopted procedure and the Monte Carlo simulations that were performed to estimate the
uncertainties of the line fluxes are given in Appendix~\ref{ap:lines}.

\subsubsection{H$\alpha$ velocity field}

\label{sec:gasvel}

The fitted positions of the strongest of all emission lines, H$\alpha$, provide the 
best estimate of the gas velocity field.
These fields, shown in  Figs.~\ref{f:g:1999dq}-\ref{f:g:07A-07R},  were analyzed with the
methods and IDL programs developed by \cite{2006MNRAS.366..787K}. The program analyzes the velocity field at several radii
and for each of them returns the inclination and position angle and quantifies the degree of deviation from pure disk rotation.
From this analysis  we also derived the redshift, the average position angle (PA) and inclination $i$ for each galaxy, which
 are listed in Table~\ref{t:hosts}.

\subsubsection{Extinction, H$\alpha$ flux, and star formation rate maps}

For the purpose of the following analysis the measured line fluxes were corrected for dust extinction
using the observed Balmer decrement $I(\mathrm{H}\alpha)/I(\mathrm{H}\beta)$ and assuming a foreground dust screen.
 For the intrinsic Balmer 
decrement $I(\mathrm{H}\alpha)/I(\mathrm{H}\beta)_{\rm intr}$ a value of 2.86 was assumed, which is appropriate for case-B 
recombination with electron temperature $T_{\rm e}=10000$\,K and electron density 10$^2$\,cm$^3$ 
\citep[e.g.,][]{2006agna.book.....O}. The dust is described by the \cite{fitzpatrick99} law with $R_V$=3.1.

The extinction-corrected H$\alpha$ flux was converted into instantaneous  SFR using the \cite{1998ARA&A..36..189K} relation:
\begin{equation}
\mathrm{SFR}\,[M_{\sun}\,\mathrm{yr}^{-1}]=7.9\times 10^{-42}\,L(\mathrm{H}\alpha),
\end{equation}
where $L$(H$\alpha$) is the H$\alpha$ luminosity in units of erg\,s$^{-1}$.

\subsubsection{ISM oxygen abundance}

The most accurate method to measure the ISM abundances -- the so-called \emph{direct} method  -- 
involves determining 
of the ionized gas electron temperature, $T_\mathrm{e}$, which is usually estimated from
 the flux ratios of auroral to nebular emission lines,
e.g. [\ion{O}{iii}]\,$\lambda\lambda$4959/5007/[\ion{O}{iii}]\,$\lambda$4363
\citep[e.g.][]{2006A&A...454L.127S,2006A&A...448..955I}. However, the
temperature-sensitive lines such as [\ion{O}{iii}]\,$\lambda$4363 are very weak and 
difficult to measure, especially in metal-rich environments. A careful
examination of our data-cubes revealed that the [\ion{O}{iii}]\,$\lambda$4363
line was not present. For this reason we used other strong emission line methods to determine the
gas oxygen abundance. Many such methods have been developed throughout the years, the
most commonly used being R$_{23}=$([\ion{O}{ii}]\,$\lambda$3727+[\ion{O}{iii}]\,
$\lambda\lambda$4959/5007)/H$\beta$ ratio-based methods
\citep{1979MNRAS.189...95P,1991ApJ...380..140M,1994ApJ...420...87Z,
2004ApJ...613..898T,2002ApJS..142...35K,
2004ApJ...617..240K,2001A&A...374..412P,2005ApJ...631..231P},
N2=log[[\ion{N}{ii}]\,$\lambda$6584/H$\alpha$]
\citep{1994ApJ...429..572S,2004MNRAS.348L..59P} and 
O3N2=log[([\ion{O}{iii}]\,
$\lambda$5007)/H$\beta$)/([\ion{N}{ii}]\,$\lambda$6584/H$\alpha$)]
\citep{1979A&A....78..200A,2004MNRAS.348L..59P}. 
More recently,
\cite{2006ApJ...652..257L,2007A&A...473..411L} and \cite{2007A&A...462..535Y}
have verified and re-calibrated these and other strong-line methods 
using Sloan Digital Sky survey (SDSS) spectroscopy. 

Unfortunately, there are large
systematic differences between the methods, which translate into a considerable uncertainty 
in the absolute metallicity scale \cite[for a recent review see, e.g.,][]{2008ApJ...681.1183K}.
 In particular, there is $\sim0.4$ dex  difference between the so-called \emph{empirical} and \emph{theoretical} strong-line methods. The  \emph{empirical} methods are calibrated against \ion{H}{II} regions and galaxies whose metallicities  
 have been previously determined by the \emph{direct} method, e.g. O3N2 and N2 \citep{2004MNRAS.348L..59P}, 
 R$_{23}-P$ \citep{2005ApJ...631..231P}. The \emph{theoretical} methods, on the other hand, are calibrated 
by matching the observed line fluxes  with those predicted by theoretical photoionization models \cite[most of the R$_{23}$-based methods, e.g.,][]{1991ApJ...380..140M,2004ApJ...617..240K,2004ApJ...613..898T,2002ApJS..142...35K}. 
The cause of these discrepancies is still not well-understood.  
Recently \cite{2010ApJS..190..233M} discussed this problem and concluded that the \emph{empirical} methods may 
underestimate the metallicity by a few tenths of dex \citep[see also][]{2007RMxAC..29...72P}, while the  \emph{theoretical} methods
overestimate it. In this situation, we followed the recommendation of \cite{2008ApJ...681.1183K} 
to use one method to compute the metallicities in all galaxies and discuss the results in \emph{relative} sense, and 
use another method to confirm the observed trends. 
As our primary method we used the \emph{empirical} O3N2 method of \cite{2004MNRAS.348L..59P} (PP04 from now on) and 
checked the results with the  \emph{theoretical} R$_{23}$ method of \cite{2004ApJ...617..240K} (KK04 from now on). 
Both methods have advantages and disadvantages, which have been discussed in several papers \citep[e.g.,][]{2008ApJ...681.1183K,2007A&A...462..535Y}.

\subsection{Stellar populations}

The star formation history and chemical evolution of a galaxy is imprinted in the  
properties of its present-day stellar populations. 
Determining the properties of the stellar populations in the galaxies  
has been a major research topic in astrophysics and through the years many different methods have been used, ranging from 
analysis of the color-magnitude diagrams \citep[CMD,][]{1972A&A....20..361F} to equivalent widths of absorption lines
 \citep[e.g., the Lick indices,][]{1994ApJS...94..687W}.
However, in most galaxies several stellar population are simultaneously present.  
Disentangling their contribution to the galaxy spectrum   
 is a very difficult task because of various astrophysical and numerical degeneracies.

\subsubsection{Full-spectrum fitting technique}
 
 Recently, the so-called \emph{evolutionary population synthesis methods}
 \citep{1968ApJ...151..547T,1996ApJS..106..307V,2003MNRAS.344.1000B} coupled with full-spectrum fitting techniques 
 \citep[e.g,][]{1999ApJ...525..144V,2001MNRAS.327..849R,2005MNRAS.358..363C,2008MNRAS.385.1998K,2009MNRAS.395...28M} have emerged as powerful means to analyze galaxy spectra. The evolutionary population synthesis methods produce synthetic galaxy spectra  
using as input theoretical evolutionary tracks, libraries of stellar spectra, initial mass function (IMF), and prescriptions for star formation and chemical evolution. The models are then compared to the observed spectra to infer the properties of the stellar populations 
that contribute to the formation of the observed spectrum.
One possible approach is to fit the observed spectrum with a linear combination of model spectra
of single stellar populations (SSP) of different ages and metallicities  \citep[e.g.,][]{2005MNRAS.358..363C,2008MNRAS.385.1998K,2009MNRAS.395...28M}.
The fitting returns the contribution of the different SSPs (called population vector) that best describe the observed spectrum, which then can be used to study the stellar populations of the galaxy. However, because of astrophysical and numerical degeneracies, and the presence of noise in the observed spectra, 
it is well-known that the solution may not be unique and the results should be interpreted with caution
\citep[e.g, see the discussion in][]{2005MNRAS.358..363C}. The best known is the age-metallicity degeneracy\footnote{Dust reddening also adds to this problem, partly because the dust extinction law may be different in different galaxies; galaxies in the Local Group are a good example for this.} where young metal-rich stellar populations are confused with older metal-poor ones \citep[see for example Fig. 10 in][]{2003MNRAS.344.1000B}. As noted by \cite{2003MNRAS.344.1000B}, while the shape of the stellar continuum is roughly the same, the strength of the metal lines increases. Therefore, analyzing well-calibrated spectra with high S/N and spectral 
resolution to resolve the absorption lines has the potential to brake the age-metallicity degeneracy.
In addition, uncertainties in the input ingredients needed for computing the SSPs, such as non-uniform coverage of the age/metallicity parameter space of the stellar libraries, IMF and the difficulties in describing some phases of the  stellar evolution  (e.g., the thermal-pulsating asymptotic giant branch phases), add even more uncertainties when interpreting the results \citep[see, e.g.,][]{2009ApJ...699..486C}.

\subsubsection{Choice of the base}

In this study we used the {\tt STARLIGHT} code described in
\cite{2005MNRAS.358..363C} and \cite{2007MNRAS.381..263A} coupled with a version of the Bruzual \& Charlot\footnote{circa 2007; unpublished}
SSP models based on the new MILES spectral library \citep{2006MNRAS.371..703S}.  The selection 
of the SSP basis is important for any full-spectrum fitting algorithm and the interpretation of the results. To minimize the computing 
time one should select few SSPs that are maximally independent and at the same time are capable of reproducing the 
 variability of the full SSP set for a given metallicity. If a large basis is selected, many of its components will be close neighbors.
This will lead to increased non-uniqueness of the solution and increase the time for the 
fitting algorithm to converge. On the other hand, if too small a basis is selected, it will not be able to capture the full 
variability of the SSP models, the fits may be poor, and the results will be unreliable. 

In our work we used the following approach to select 
the basis. For a given metallicity all SSPs were 
normalized to the flux in the 4600-4800\,\AA\ interval. Then the evolution of the flux in seven spectral windows in the range 3700-7000\,\AA\ was tracked as a function of the SSP age. The goal was to identify age intervals where the flux in \emph{all} seven spectral 
windows evolves linearly (or close to) with time. If such intervals exist, then the SSPs within them are not independent; all SSPs in a given interval can be closely reproduced as a linear combination of the two SSPs at the extremes. By selecting the basis at the ages connecting the 
linear intervals we form a small independent set of basis vectors, which at the same time can reproduce the SSPs at all other ages. 
Following this approach we were able to 
select $N_{\ast}=16$ or 17 SSPs per metallicity that formed our fitting basis of 66 SSP models with ages between 1 Myr and 18 Gyr, 
and four metallicities $Z$=0.004, 0.008, 0.02 (the solar metallicity) and 0.05.

\subsubsection{Voronoi binning}

To increase the S/N in the outer parts of the galaxies the data cubes were spatially binned using adaptive Voronoi tessellations \citep{2003MNRAS.342..345C,2006MNRAS.368..497D}. 
The binning of the spaxels was determined from the S/N measured in the interval 4580-4640\,\AA, after discarding the spectra with S/N$<$1. The targeted S/N of the binned spaxels was S/N$\sim$20, with the exception of the host of \object{SN~2006te}, for which a lower S/N of 15 was used. To keep the spatial resolution reasonably small, an upper
limit of the size of the bins was also imposed: 5 for \object{NGC976}, 17 for \object{NGC495}, and 12 for the remaining four.

\begin{table*}[!th]
\caption{Total galaxy SFR, SFR surface density $\Sigma_{\rm SFR}$ and gas extinction $A_{\rm V}$
 at the SN location derived from our observations.  } 
\label{t:sfr}
\begin{tabular}{@{}llcccccc@{}}
\hline
\hline\noalign{\smallskip}
SN   	&    Host galaxy &   \multicolumn{4}{c}{total SFR\,[M$_{\sun}$\,yr$^{-1}$] } &  $\Sigma_{\rm SFR}$ at SN position & $A_{\rm V}$ at SN position  \\
\cline{3-6}\noalign{\smallskip}
	 &     	&  H$\alpha$\tablefootmark{a} & \multicolumn{2}{c}{{\tt STARLIGHT}\tablefootmark{a}} & Neill et al.\tablefootmark{b} &  [M$_{\sun}$\,yr$^{-1}$\,kpc$^{-2}$] &   [mag] \\
	 \cline{4-5}\noalign{\smallskip}
	 &     	&   & $<$50 Myr & $<$0.5 Gyr &  &   &    \\
	 
\hline\noalign{\smallskip}
1999dq 	&  NGC 0976    	 &  5.30    (0.06) & 6.8 &16.0 &  10.4 (2.5,36.6)  & 8.9(1.2)$\times$10$^{-2}$    &   0.97 (0.18) \\                                                                                                                  
2001fe 	&  UGC 5129      &  0.76    (0.02) & 1.0 & 3.6 &	3.2 (0.6,10.4) & 1.5(0.4)$\times$10$^{-2}$    &   0.79 (0.35) \\ 
2006te 	&  CGCG 207-042  &  2.03    (0.05) & 1.6 & 5.2 &	2.9 (1.3,6.9)  & 8.3(2.2)$\times$10$^{-3}$    &   0.74 (0.33) \\ 
2007A  	&  NGC 105 NED02 &  3.42    (0.04) & 5.3 & 9.0 &   11.6 (2.4,27.5) & 2.6(0.4)$\times$10$^{-2}$    &   0.52 (0.22) \\ 
1997cw 	&  NGC 105 NED02 &  3.42    (0.04) & 5.3 & 9.0 &   11.6 (2.4,27.5) & 1.6(0.3)$\times$10$^{-2}$    &   0.06 (0.22) \\ 
2007R  	&  UGC 4008 NED01&  5.33    (0.12) & 5.8 &24.1 &	3.6 (2.1,205.6)& 3.2(0.8)$\times$10$^{-2}$    &   1.17 (0.32) \\ 
\hline 
\end{tabular}\\
\tablefoottext{a}{this work;}
\tablefoottext{b}{from \cite{2009ApJ...707.1449N}. The errors are asymmetric and the values in the parentheses are the $\mp1\sigma$ uncertainties.}
\end{table*}

\subsubsection{{\tt STARLIGHT} fits}

\label{sec:starfits}

The Voronoi-binned spectra along with the un-binned ones were fitted with the 
{\tt STARLIGHT} code allowing for \emph{all}  SSPs to be reddened by the same amount of dust described by the \cite{ext_law} law. 
In our analysis, the spectra and the basis were normalized to the mean flux in the region 4580-4620\AA. Thus the population 
vector  is the fractional contribution $x_j$ of the different SSP models at $\sim$4600\AA.
In addition to the population vector the code also returns the fractional contributions $\mu_j$ of each 
SSP to the total stellar mass of the galaxy, which is the more relevant  physical quantity. The code also 
returns the velocity shift and the Gaussian broadening that need to be applied to the model in order to fit the 
observed spectrum. The shifts provide the velocity maps for the stars and the broadening is 
related to the velocity dispersion of the stars. From the population vectors we can compute the mass- and 
light-weighted mean age and metallicity following \cite{2005MNRAS.358..363C}:
\begin{equation}
\langle \log t_\ast\rangle_{\rm L/M}=\sum\limits_{j=1}^{N_{\ast}} w_j\,\log t_j
\end{equation}
\begin{equation}
\langle Z_\ast\rangle_{\rm L/M}=\sum\limits_{j=1}^{N_{\ast}} w_j\,Z_j,
\end{equation}
where $t_j$ and $Z_j$ are the age and the metallicity of the $j$-th SSP model, and  $w_j=x_j$ or $w_j=\mu_j$ for light- and mass-weighted quantities, respectively.

\subsubsection{Compressed population vectors}

The simulations performed by \cite{2005MNRAS.358..363C} demonstrated that the individual components of the population 
vectors computed by {\tt STARLIGHT} are very uncertain. Instead of analyzing  the individual components, \cite{2005MNRAS.358..363C} showed that a coarsely binned version of the population vectors provides a more robust description of the current stellar content of the galaxies. Thus, following \cite{2004MNRAS.355..273C} and \cite{2005MNRAS.358..363C}, we formed compressed the population vectors in three age bins: young (age $<$ 300 Myr), intermediate (300 Myr $<$ age $<$ 2.4 Gyr), and old (age $>$ 2.4 Gyr) stellar populations.

\section{Results}

\subsection{Ionized gas}

Figures~\ref{f:bpt}-\ref{f:rad_met} show the main results obtained from the analysis of the emission line fluxes.
The 2D maps of the galaxy properties that are discussed in this section are shown by galaxy 
in Figs~\ref{f:g:1999dq}-\ref{f:g:07A-07R}.

\subsubsection{BPT diagnostic diagram}

\begin{figure}[!h]
\includegraphics [width=8.8cm]{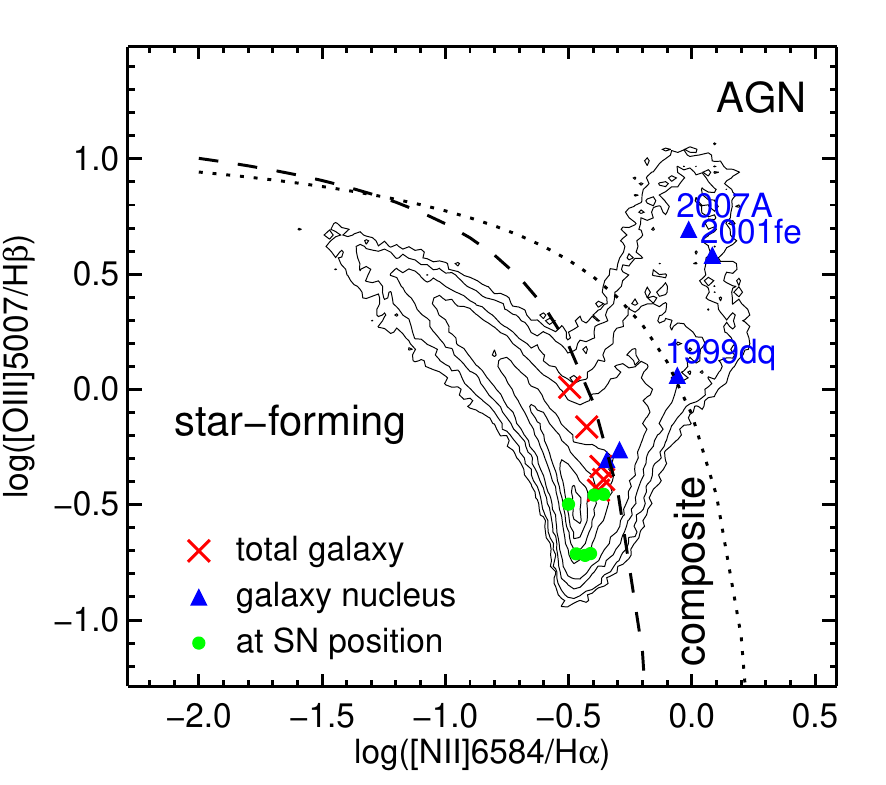} 
\caption{BPT diagram \citep{1981PASP...93....5B}. The contours show the density of
SDSS emission line galaxies. The dotted and the dashed lines of \cite{2001ApJ...556..121K} and 
\cite{2003MNRAS.346.1055K}, respectively, separate star-forming galaxies, AGNs, and composite galaxies.
The filled circles, filled triangles, and crosses show the measurements at the position of the SN, the galaxy nucleus, and the total 
galaxy spectrum, respectively. }
\label{f:bpt}
\end{figure}

 \cite{1981PASP...93....5B} introduced several diagnostic diagrams to segregate spectra 
of emission-line galaxies and AGNs according to their main excitation mechanism. 
These diagrams are based on easily measured optical emission line flux ratios.
Figure\,\ref{f:bpt}
shows the positions of the galaxies in our sample on the log([\ion{N}{ii}]\,$\lambda$6584/H$\alpha$) -- log([\ion{O}{iii}]\,
$\lambda$5007/H$\beta$) diagnostic diagram. The filled circles, filled triangles, 
and crosses show the measurements at the position of the SN, the galaxy nucleus, and the total 
galaxy spectrum, respectively. The dotted and dashed lines show two widely used criteria to separate 
emission-line galaxies and AGNs introduced by \cite{2001ApJ...556..121K} and \cite{2003MNRAS.346.1055K}, respectively.
From Fig.\,\ref{f:bpt} it is evident that the hosts of SNe 2001fe and 2007A/1997cw harbor AGNs and the host of \object{SN~199dq}  is on the border 
of composite galaxies and AGNs.
However, the line ratios measured in the total spectra of these three galaxies still
fall into the  star-forming region of the BPT diagram, which suggests that
the AGNs are not strong enough to significantly affect the total galaxy spectra.
At high redshift good S/N, spatially resolved spectroscopy is 
difficult to obtain and weak AGNs may remain unrecognized in slit spectroscopy because typically the 
whole galaxy falls into the slit. The presence of AGNs, even though  weak, may still bias the metallicity estimation 
from integrated galaxy spectra.

The locations of the SNe fall into the (small) region of the BPT diagram with the highest 
density of SDSS galaxies. This is the region where the metal-rich 
galaxies are typically found \citep[see, e.g.,][]{2007MNRAS.375L..16C}. High metallicity 
in this region of the BPT diagram is also expected from the O3N2 method 
 for metallicity estimation \citep{1979A&A....78..200A,2004MNRAS.348L..59P}. 
These are indications that for
the emission line galaxies in our sample, the SNe likely exploded in metal-rich environments.

\subsubsection{H$\alpha$ velocity field}

\begin{figure*}[!t]
\centering

\includegraphics [width=18cm]{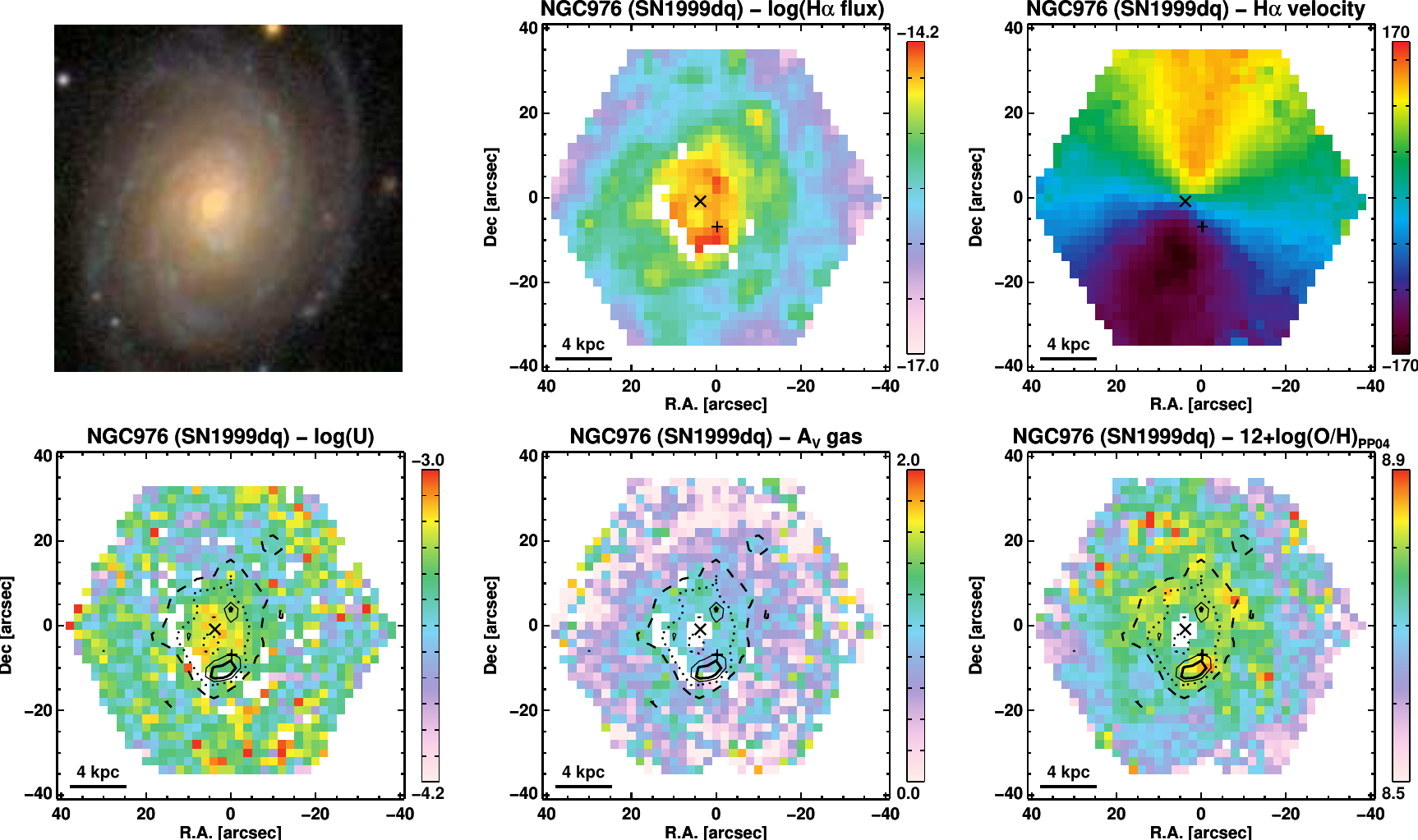}

\caption{{\bf Upper row:} From left to right, the color SDSS image of \object{NGC~976}, the observed H$\alpha$ flux and velocity maps. {\bf Lower row:} The ionization parameter $\log(U)$,  the visual extinction $A_{\rm V}$ estimated from the Balmer decrement and the metallicity map derived with the \cite{2004MNRAS.348L..59P} O3N2 method.  In all  
maps presented in this paper,  $\times$ marks the galaxy center and $+$ the SN position. The four contour levels  overplotted on the extinction and metallicity maps are derived 
from the H$\alpha$ map. The four levels are 0.8, 0.6, 0.4 and 0.2 of the maximum H$\alpha$ flux. The $x,y$ 
coordinates are in arcsec with respect to the map centers. The orientation of the images is north -- up, east -- left.}
\label{f:g:1999dq}
\end{figure*}

\begin{figure*}[!ht]
\centering

\includegraphics [width=18cm]{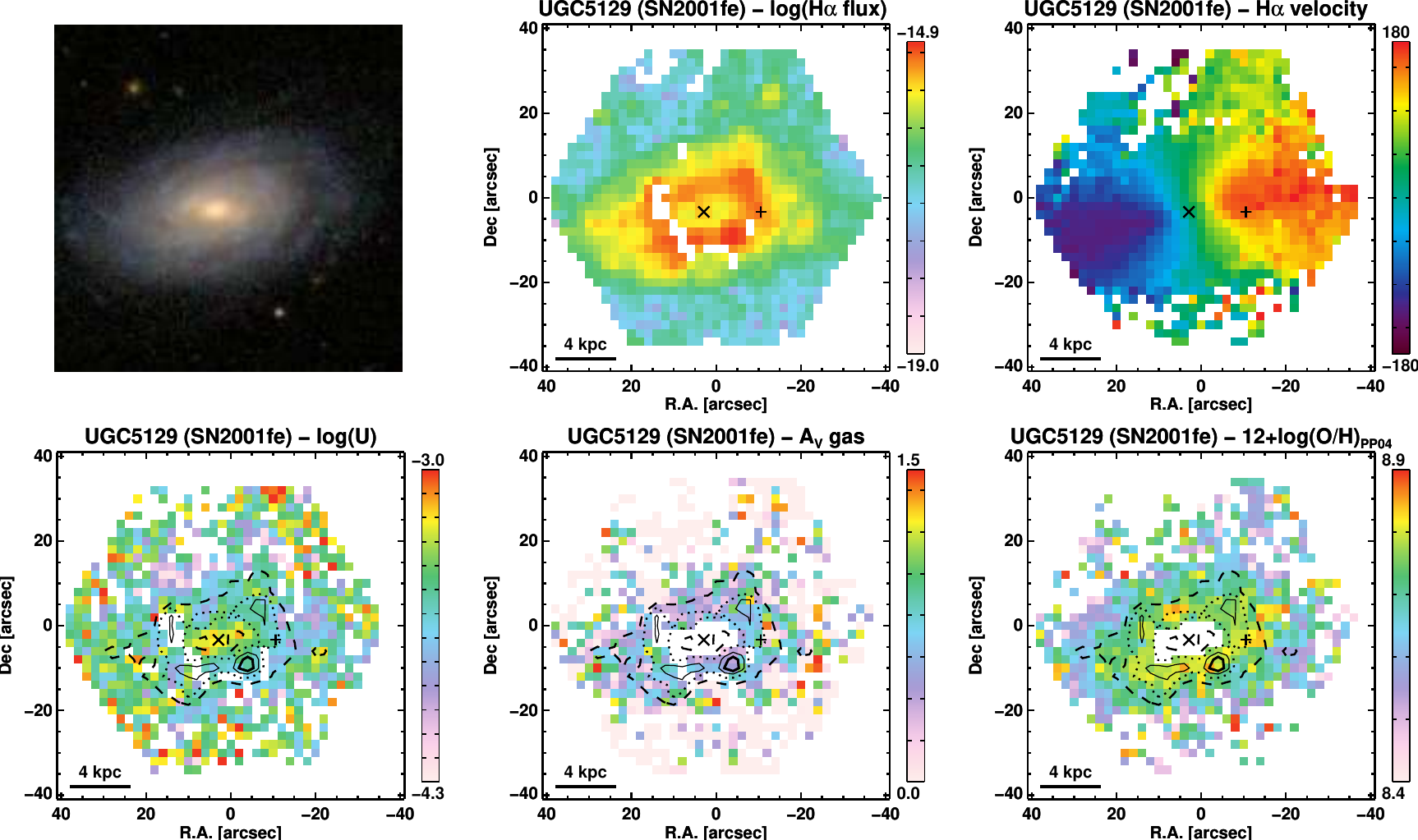}\vspace{1cm} \\
\includegraphics [width=18cm]{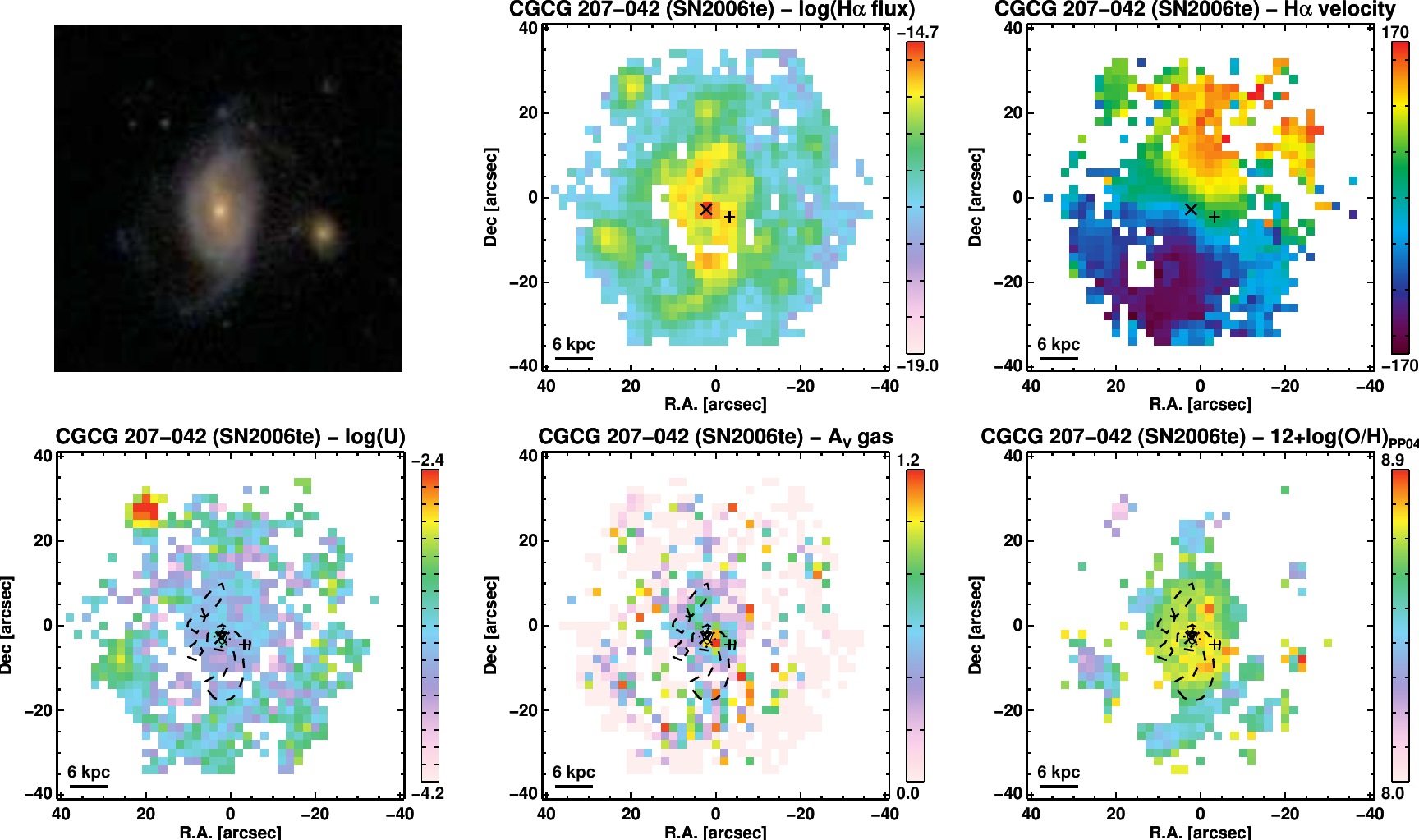}

\caption{Same as Fig.~\ref{f:g:1999dq} but for \object{UGC~5129} and \object{CGCG~207-042}.}
\label{f:g:01fe-06te}
\end{figure*}

\begin{figure*}[!ht]
\centering

\includegraphics [width=18cm]{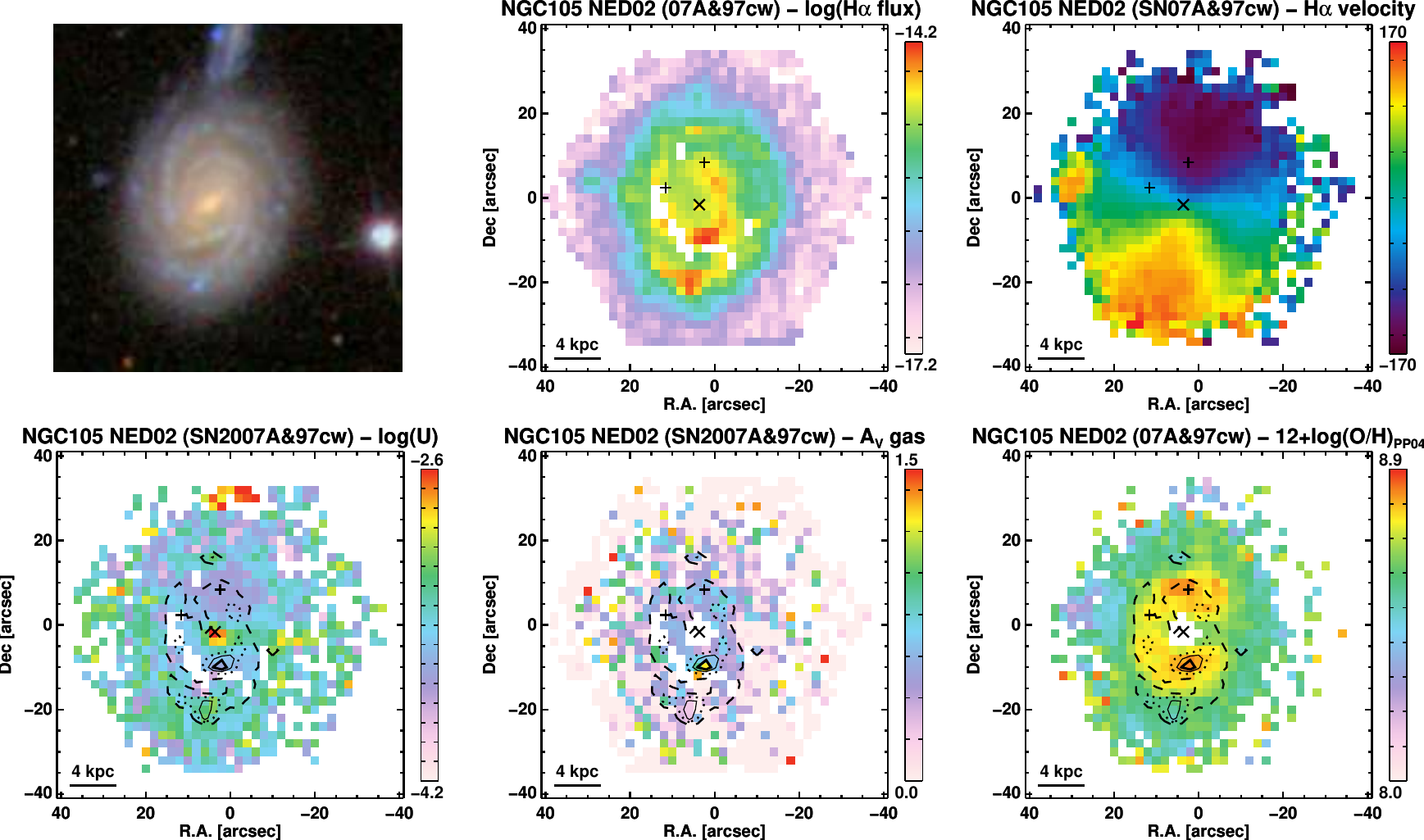}\vspace{1cm} \\
\includegraphics [width=18cm]{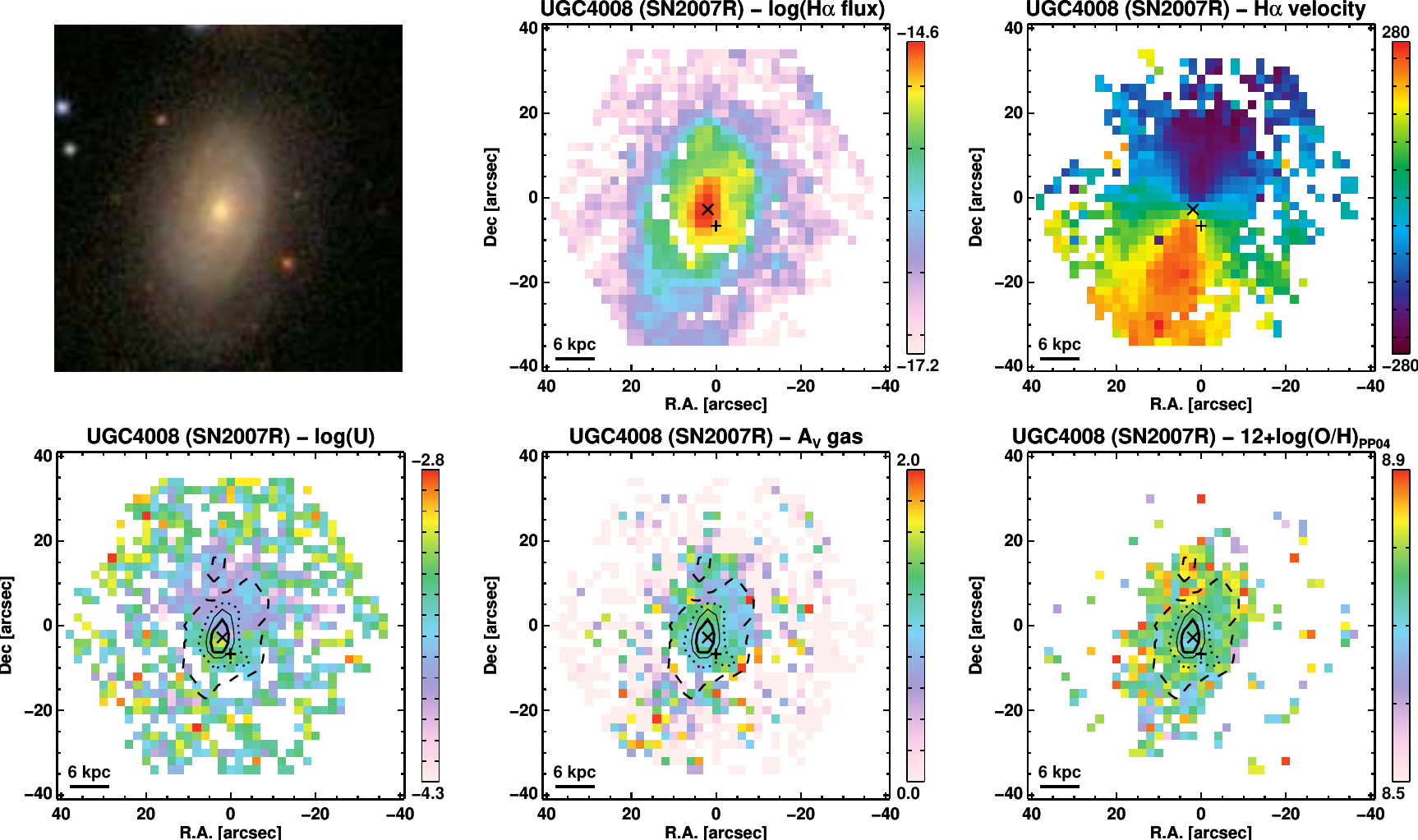}

\caption{Same as Fig.~\ref{f:g:1999dq} but for \object{NGC~105 NED02} and \object{UGC~4008 NED01}.}
\label{f:g:07A-07R}
\end{figure*}

The gas velocity maps derived from the H$\alpha$ emission line
show smooth gradients and no apparent irregularities.
The analysis with the method of \cite{2006MNRAS.366..787K} shows that the velocity fields of all five galaxies 
are consistent with pure disk rotation 
without signs of significant disturbances. The $\sigma$-maps (not shown here) derived form the width of H$\alpha$ emission line 
also show a simple structure with a single peak at the center. These results suggest that the galaxies in our sample are likely relaxed systems.

\subsubsection{H$\alpha$ flux, extinction, and star formation rate}
\label{sec:ha}

 On small scales the H$\alpha$ flux distribution follows the spiral arms visible in the broad-band images.
Overall, the H$\alpha$ flux increases toward the centers of the galaxies. It can be seen that all six SNe 
 in these galaxies are projected onto regions with strong 
H$\alpha$ emission with fluxes above the galaxy-average. 
In the late-type spirals \object{NGC~976}, \object{UGC~5129}, and \object{NGC~105 NED02} there is a clear H$\alpha$ flux 
deficit in the bulge. Interestingly, these are also the galaxies that have AGNs. However, these two phenomena are 
probably unrelated. \cite{2009A&A...501..207J} studied the radial distribution of H$\alpha$ emission 
in a large sample of spiral galaxies and found that the late-type spirals (Sc+) show  a H$\alpha$ flux deficit in the 
bulge regardless of the presence of bars. This effect is much less pronounced in the Sa-type spirals or 
even absent in their barred counterparts, which tend to have a high concentration of the H$\alpha$ emission 
toward the center. This is clearly the case for \object{UGC~4008 NED01}, which
 is the only Sa emission line galaxy in our sample.

According to \cite{2009A&A...501..207J}, there is a significant difference in the H$\alpha$ radial profile of barred and unbarred Sb galaxies. 
The unbarred Sb galaxies show a smooth profile similar to Sa galaxies. The barred counterparts
have a strong peak of  H$\alpha$ emission at the centers, followed by a decrease of the H$\alpha$ flux, before it increases again
because of the  H$\alpha$ emission ring at the outer radius of the bar. 
The barred Sb galaxy \object{CGCG~207-042} shows exactly the same characteristics with a clear H$\alpha$  emission ring at the 
outer radius of the bar. Thus, the radial H$\alpha$ emission profiles in our galaxies are consistent with the findings of
 \cite{2009A&A...501..207J}.

The gas extinction maps presented in Figs~\ref{f:g:1999dq}-\ref{f:g:07A-07R} also show increase of the extinction toward the galaxy center. Although the extinction
maps do not show small-scale structures as clearly as in the  H$\alpha$ flux maps, there is a general trend that 
the extinction increases with the H$\alpha$ flux. This is expected because an increased amount of dust is typically observed 
in the regions of active star formation.
The total extinction along the lines of sights of the SN position is low, except for SN 1997cw (marked with the leftmost of the three signs). 
The extinction along the SN lines-of-sight is given in Table\,\ref{t:sfr}.

Table\,\ref{t:sfr} lists the total on-going SFR and the SFR surface density, 
$\Sigma_{\rm SFR}$, at the SN positions derived from the extinction-corrected H$\alpha$ flux map. The $\Sigma_{\rm SFR}$ 
values at the SN position are consistent with the disk-averaged values for normal spiral galaxies 
 \cite[see Fig.~5 in][]{1998ARA&A..36..189K}. Our values fall in the upper half of the \cite{1998ARA&A..36..189K}  distribution, 
which can be attributed to the SN being projected on regions with higher-than-average H$\alpha$ flux. 

For comparison the total SFR and its confidence intervals derived by \cite{2009ApJ...707.1449N} are also given in Table\,\ref{t:sfr}. 
 In general, our values are consistent with  \cite{2009ApJ...707.1449N}, although in all cases but one we derive lower values. 
  However, it should be noted that the values of \cite{2009ApJ...707.1449N} were derived 
 with a completely different technique -- fitting model galaxy SEDs to broad-band photometry -- 
 and represent the average SFR during the last 0.5 Gyr, while our estimates from the  H$\alpha$ flux 
 represent the very resent, $<20$ Myr, SFR.

\subsubsection{Ionization parameter and electron density}

The  ionization parameter $\log(U)$ -- the ratio of the ionizing photon
density to the gas density -- is a measure of the degree of
ionization of the nebula and can be determined from the ratio of two
lines of the same element corresponding to two different
ionization states.  The ionization maps of the galaxies were computed
from the ratio of the [\ion{O}{ii}]\,$\lambda$3727 and
[\ion{O}{iii}]\,$\lambda$5007 lines using the relation of \cite{2000MNRAS.318..462D}.

The three AGN galaxies clearly show an increased degree of ionization  
toward the center, while the remaining two galaxies do not. In three of the galaxies,
 \object{NGC~976}, \object{CGCG 207-042}, and \object{NGC~105 NED02}, there is also a hint of
increasing of the ionization parameter toward the outer spirals. The mean ionization parameters for 
all five galaxies fall into a rather narrow interval of $\log(U)= -3.6\div-3.4$. The [\ion{O}{ii}]\,$\lambda$3727/
[\ion{O}{iii}]\,$\lambda$5007 line ratio is known to provide lower values for the ionization parameters compared to 
other available methods \cite[e.g.,][]{2000MNRAS.318..462D}. In comparison with the ionization parameter
maps that were computed as part of the \cite{2004ApJ...617..240K} method for oxygen abundance determinations, the [\ion{O}{ii}]\,$\lambda$3727/
[\ion{O}{iii}]\,$\lambda$5007-based maps show very similar features, but are shifted toward lower values by $\sim0.3-0.4$~dex. Even taking this offset into account, the average values for our galaxies fall into the lower end of the distribution of the \ion{H}{ii} galaxies studied by 
\cite{1998Ap&SS.263..143D}. This part of the distribution is mostly populated with  \ion{H}{ii} galaxies without measurable [\ion{O}{iii}]\,$\lambda$4363 line, which tend to be metal-rich.

Given the spectral resolution and wavelength range of our spectroscopy, the electron density, $n_\mathrm{e}$, can be estimated 
only from the flux ratio of the [\ion{S}{ii}]\,$\lambda\lambda$6716/6731 lines  \citep{2006agna.book.....O}. 
The ratio of these two lines is sensitive to $n_\mathrm{e}$ in the range $\sim10^2-10^4$~cm$^{-3}$. 
Unfortunately, for the two highest redshift galaxies in our
sample, \object{CGCG 207-042} and \object{UGC 4008 NED01}, the [\ion{S}{ii}] lines are outside the covered wavelength range. For the
remaining three galaxies the ratio is constant $\sim$1.4 across the galaxy and no apparent structure is visible. 
Ratios of $\sim$1.4 indicate low electron density $n_\mathrm{e}\leq10^2$~cm$^{-3}$ and are similar to the measurements of $n_\mathrm{e}$ in other galaxies,
 e.g. the sample of SDSS galaxies studied by \cite{2004ApJS..153..429K}. The low electron density suggests that there are no shocks in the galaxy.

\subsubsection{ISM oxygen abundance}

Figures~\ref{f:g:1999dq}-\ref{f:g:07A-07R}  also show
 the distributions of the metallicity (indicated as 12+log(O/H)$_{\rm PP04}$) estimated with the 
O3N2 method of \cite{2004MNRAS.348L..59P}. For the three galaxies that likely harbor AGNs, namely, \object{NGC~976}, \object{UGC~5129}, and
 \object{NGC~105~NED02}, the central regions affected by the AGN are masked. The decrease of the S/N in the outer parts of the galaxies
affects the measurements of the line ratios and some spaxels are also marked as affected by AGN according to  the \cite{2001ApJ...556..121K} 
 criterion. Because no AGN activity is expected in the outer parts of the galaxies the line ratios measured in those spaxels are likely 
 dominated by the noise and have also been masked in the plots.
  In these three galaxies there are indications for ring-like structures with enhanced metallicity and  H$\alpha$ flux. However,
  it is difficult to assess whether these are real structures or artifacts caused by the central AGN altering the line ratios. 
  Thus, these ring-like structures should be regarded with caution. 

\begin{figure}[!ht]
\includegraphics [width=8cm]{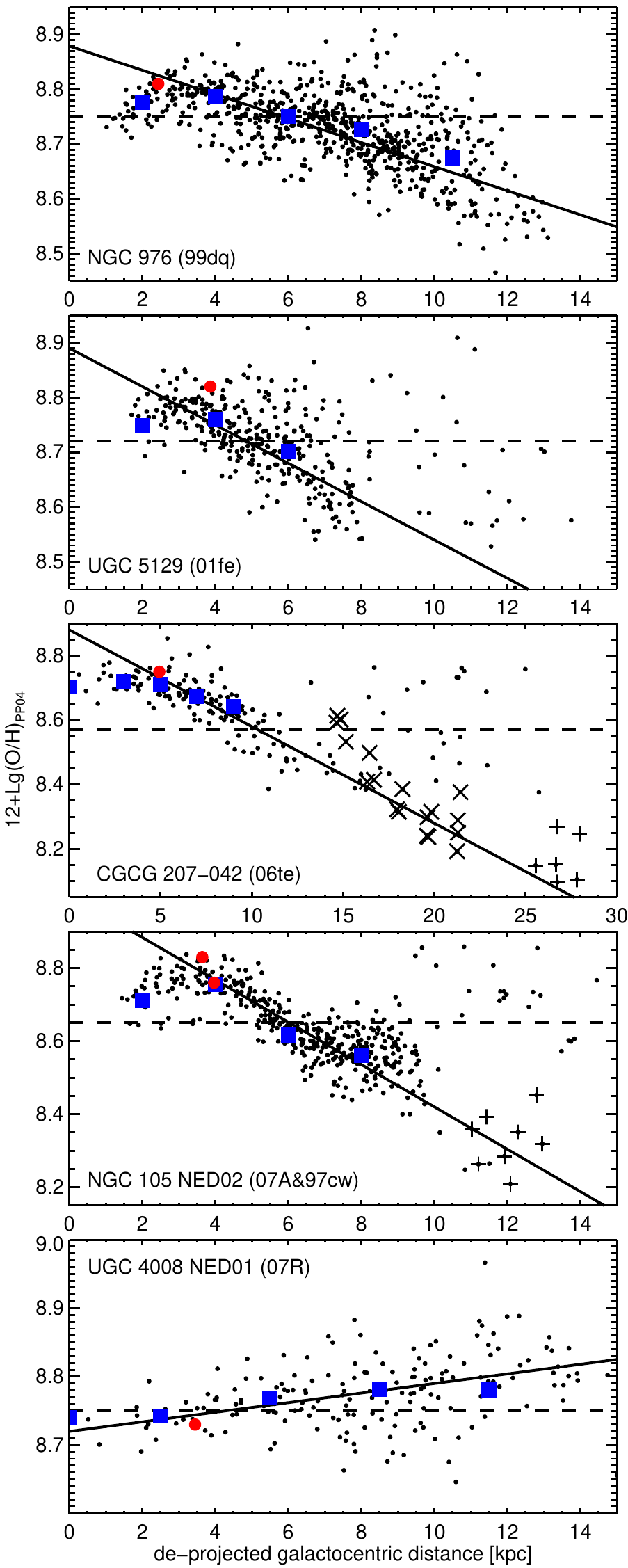} 
\caption{Radial dependence of 12+log(O/H)$_{\rm PP04}$. The red dots show the measurement at the SN position. The horizontal dashed lines
indicate the metallicity measured from the total galaxy spectra (Table~\ref{t:oh}).
As a guide to the eye, we plot metallicity gradients of 
$-0.022$, $-0.035$,$-0.030$, $-0.058$ and $+0.007$ dex\,kpc$^{-1}$ (from top to bottom) with the solid lines. 
We estimate that the accuracy of these gradients is $\sim0.005$.
The blue squares show the metallicities measured from the azimuthally 
averaged spectra.
The points shown with the cross and plus signs are the metallicities of the \ion{H}{ii} regions 
discussed in Sect.~\ref{notes:gas}.}
\label{f:rad_met}
\end{figure}

Figure~\ref{f:rad_met} shows the dependence of the metallicity on the de-projected galactocentric distance, which was computed from the position angle and inclination derived from the analysis of the H$\alpha$ velocity fields.
Excluding the host of \object{SN~2007R}, which has a quite uniform metallicity distribution, the other four galaxies 
show decreasing of the metallicity with the radius. 
The plots also suggest that the high metallicities measured in some of the outermost spaxels in  \object{UGC~5129}, \object{CGCG 207-042}, and  \object{NGC~105~NED02} are most likely due to noise in the line flux measurements.
The solid lines are guides to the eye, showing  metallicity gradients of 
$-0.022$, $-0.035$,$-0.030$ and $-0.058$ dex\,kpc$^{-1}$ for \object{NGC~976}, \object{CGCG 207-042}, \object{UGC~5129}, and
 \object{NGC~105~NED02}, respecively. These metallicity gradients fall well into the range observed in other nearby galaxies, e.g. \object{M51}
\citep{2004ApJ...615..228B}, \object{NGC~300} \citep{2009ApJ...700..309B}, and \object{NGC~628}
\citep{2011MNRAS.415.2439R}. In contrast, in \object{UGC 4008 NED01} the metatllicity is nearly uniformly distributed with a hint of a very small positive gradient of $+0.007$ dex\,kpc$^{-1}$. 
The blue squares in Fig.~\ref{f:rad_met} show the metalliciy estimates from the azimutally averaged spectra described in Sect.~\ref{sec:analysis}. 
These estimates trace  the measurements on the individual spaxel spectra very well.

Figure~\ref{f:rad_met} and the inspection of the 
2D maps reveals that the SN explosion sites are projected onto regions that have the highest, or close to the highest,
 metallicity within the corresponding galaxy.  Table\,\ref{t:oh} shows
the metallicity measurements in the total galaxy spectra, at the nucleus, and at the SN position. The metallicities at the SN positions in all five galaxies are very similar to each other, 12+log(O/H)$_{\rm PP04}\sim8.8$ and 12+log(O/H)$_{\rm KK04}\sim9.1$, and are on average by 0.1 dex higher than the metallicities measured from the total galaxy spectra. For the three galaxies that host AGNs we also computed the metallicity in the total spectra, excluding the central spaxels, which are affected by the AGNs. The metallicity was in all cases identical to the one measured from the total spectra that included the central spaxels (the latter are given in Table\,\ref{t:oh}). This result shows that in these cases the AGNs are too weak to significantly affect the total galaxy spectrum and the metallicity estimation.

\begin{table*}
\caption{ISM metallicity estimates from the total galaxy spectra, at the nucleus, and at the position of the SN using the PP04 and KK04 
methods.} 
\label{t:oh}
\begin{tabular}{@{}llcccccccc@{}}
\hline
\hline\noalign{\smallskip}
SN   	&    Host galaxy&   \multicolumn{8}{c}{12+log(O/H)}   \\
\cline{3-10}
	 &      	&  \multicolumn{2}{c}{total galaxy spectrum}  &   &   \multicolumn{2}{c}{nucleus}   & &  \multicolumn{2}{c}{at  the SN position}    \\
 \cline{3-4}\cline{6-7}\cline{9-10}\noalign{\smallskip}
 	    &      	        & PP04  & KK04   &   &  PP04  & KK04  &     &PP04  & KK04 \\ 
\hline\noalign{\smallskip}
1999dq 	    &  NGC 976    	 &  8.75   (0.17) & 9.08   (0.05)  &   &  8.69\tablefootmark{a} (0.08) &  8.84\tablefootmark{a}   (0.23)  &     &   8.81   (0.05) & 9.15   (0.03)   \\ 
2001fe\tablefootmark{b} 	&  UGC 5129      &  8.72   (0.16) & 8.98   (0.11)  &   &  $\dots$      &  $\dots$        &     &   8.82   (0.10) & 9.10   (0.09)   \\ 
2006te 	    &  CGCG 207-042  &  8.57   (0.14) & 8.96   (0.10)  &   &  8.72	(0.06) &  9.10   (0.05)  &     &   8.75   (0.09) & 9.03   (0.13)   \\ 
2007A\tablefootmark{b} 	&  NGC 105 NED02 &  8.65   (0.14) & 9.02   (0.08)  &   &  $\dots$      &  $\dots$        &     &   8.83   (0.06) & 9.09   (0.07)   \\ 
1997cw\tablefootmark{b} 	&  NGC 105 NED02 &  8.65   (0.14) & 9.02   (0.08)  &   &  $\dots$      &  $\dots$        &     &   8.76   (0.06) & 9.10   (0.04)   \\ 
2007R  	    &  UGC 4008 NED01&  8.75   (0.16) & 9.08   (0.06)  &   &  8.73	(0.12) &  8.66   (0.55)  &     &   8.73   (0.09) & 9.12   (0.06)   \\ 
\hline 
\end{tabular}\\
\tablefoottext{a}{may have AGN contamination.}
\tablefoottext{b}{the metallicity not measured at the center because these galaxies harbor AGNs.}

\end{table*}

\subsubsection{Notes on the individual galaxies}
\label{notes:gas}

\noindent
\object{{\bf NGC 976 }}: The metallicity distribution is nearly symmetric around the galaxy center except for a somewhat extended region 
located at coordinates (+8,+22). The metallicity of this region is higher compared to the other parts of the galaxy at the 
same radial distance. Examining the 
$[\ion{O}{iii}]5007/{\rm H}\beta$ and $[\ion{N}{ii}]6584/{\rm H}\alpha$ maps (not shown in the paper), we noted that
this is caused by an asymmetry in $[\ion{O}{iii}]5007/{\rm H}\beta$. Both ratios are nearly symmetrically distributed 
around the galaxy center except for the region at (+8,+22), which has a lower $[\ion{O}{iii}]5007/{\rm H}\beta$ ratio
resulting in a higher metallicity estimate. 
There is also a slight decrease of the degree of  ionization at the same location. The galaxy was included by 
\cite{1997ApJ...485..552M} in the control sample for their study of the cause of the elevated star formation 
in Seyfert 2 compared with Seyfert 1 galaxies. The authors found no obvious trigger of star formation in \object{NGC 976}.

\noindent
\object{{\bf  NGC 495}}: This red barred Sa galaxy shows no emission lines. \cite{2002AJ....123.3018M} found it to be a member of a 
 poor galaxy cluster, which was  the richest cluster among those studied in their work, however.  It is therefore possible  that 
the gas component of \object{NGC 495} was separated from the galaxy by the tidal interaction with the other cluster members.

\noindent
\object{{\bf  UGC 5129}}: This galaxy was included in the study of isolated disk galaxies by \cite{2004A&A...420..873V}. It was
included in the final list of 203 galaxies (out of an initial 1706) that were likely not affected by other galaxies during the last
few Gyr of their evolution.

\noindent
\object{{\bf CGCG~207-042}}: The spirals arms of the galaxy are barely visible in the SDSS image. 
However, there are three \ion{H}{ii} regions along one of them
that are clearly visible in the H$\alpha$ map. They are roughly located at ($x,y$)  
coordinates (+8,--27), (+27,--10) and (+20,+27). The metallicity decreases considerably along the spiral arm, which is also 
accompanied by a strong increase of the ionization parameter. The two outermost \ion{H}{ii} regions are shown with 
blue and magenta points in Fig.~\ref{f:rad_met}.

\noindent
\object{{\bf NGC~105~NED02}}: 
In the H$\alpha$ velocity map there is a spot located at (+28,+4) that clearly does not follow the velocity
of the underlying part of the galaxy
but moves away $\sim120$ km\,s$^{-1}$ faster. At the same position there is a very faint spot in 
the broad-band images. This sport also clearly shows increased H$\alpha$ emission and a marginal increase of the ionization parameter. 
The metallicity of the spot is lower than the rest of the galaxy by at least 0.2 dex. The points corresponding to this 
spot are shown with blue points in Fig.~\ref{f:rad_met}. Given the properties of this feature, it is possible that this is a dwarf satellite 
galaxy of \object{NGC~105~NED02}. Another interesting feature is that the central ring-like pattern of increased metallicity 
is interrupted by a region of slightly lower metallicity located at coordinates (--3,--2). Examining the 
$[\ion{O}{iii}]5007/{\rm H}\beta$ and $[\ion{N}{ii}]6584/{\rm H}\alpha$ maps, we again noted that
this is caused by an asymmetry in $[\ion{O}{iii}]5007/{\rm H}\beta$. The  $[\ion{O}{iii}]5007/{\rm H}\beta$ ratio  at (--3,--2) is slightly higher and causes the lower metallicity estimate.

\noindent
\object{{\bf UGC~4008~NED01}}: This is the only galaxy in our sample that shows a positive metallicity gradient.

\subsection{Stellar populations}

\subsubsection{Stellar \emph{vs.} gas dust extinction}

{\tt STARLIGHT} fits provide an estimate of the extinction by dust suffered by the stellar light. The assumption that the stellar populations
of different age are subject to the same extinction is probably not entirely correct. It is reasonable to assume that the young populations
can be still embedded in the dusty nebula where the stars formed and can be subject to higher extinction. {\tt STARLIGHT} has the capability 
to take this into account and can determine different extinctions for the different SSP models. However, this approach adds additional uncertainty to 
the already complex problem of recovering the properties of the stellar populations. We have chosen to assume a single extinction for all SSPs.

The extinction maps of the star light are shown in Fig.~\ref{f:popaveL} and there were no easily identifiable features in them. In comparison
with the extinction derived from the emission lines (Figs~\ref{f:g:1999dq}-\ref{f:g:07A-07R}), the extinction derived by {\tt STARLIGHT} fits is lower. The relation 
between the star and gas extinction shows considerable scatter and the two quantities appear to be uncorrelated, except for \object{NGC~105~NED02}.
In this galaxy there is a clear linear relation between the star and gas extinction, with the gas extinction being about twice the star's extinction.
A similar relation was also derived by \cite{2005MNRAS.358..363C} in their analysis of a sample of SDSS galaxies.

\onlfig{7}{
\begin{figure*}[!ht]
\centering

\includegraphics [width=17cm]{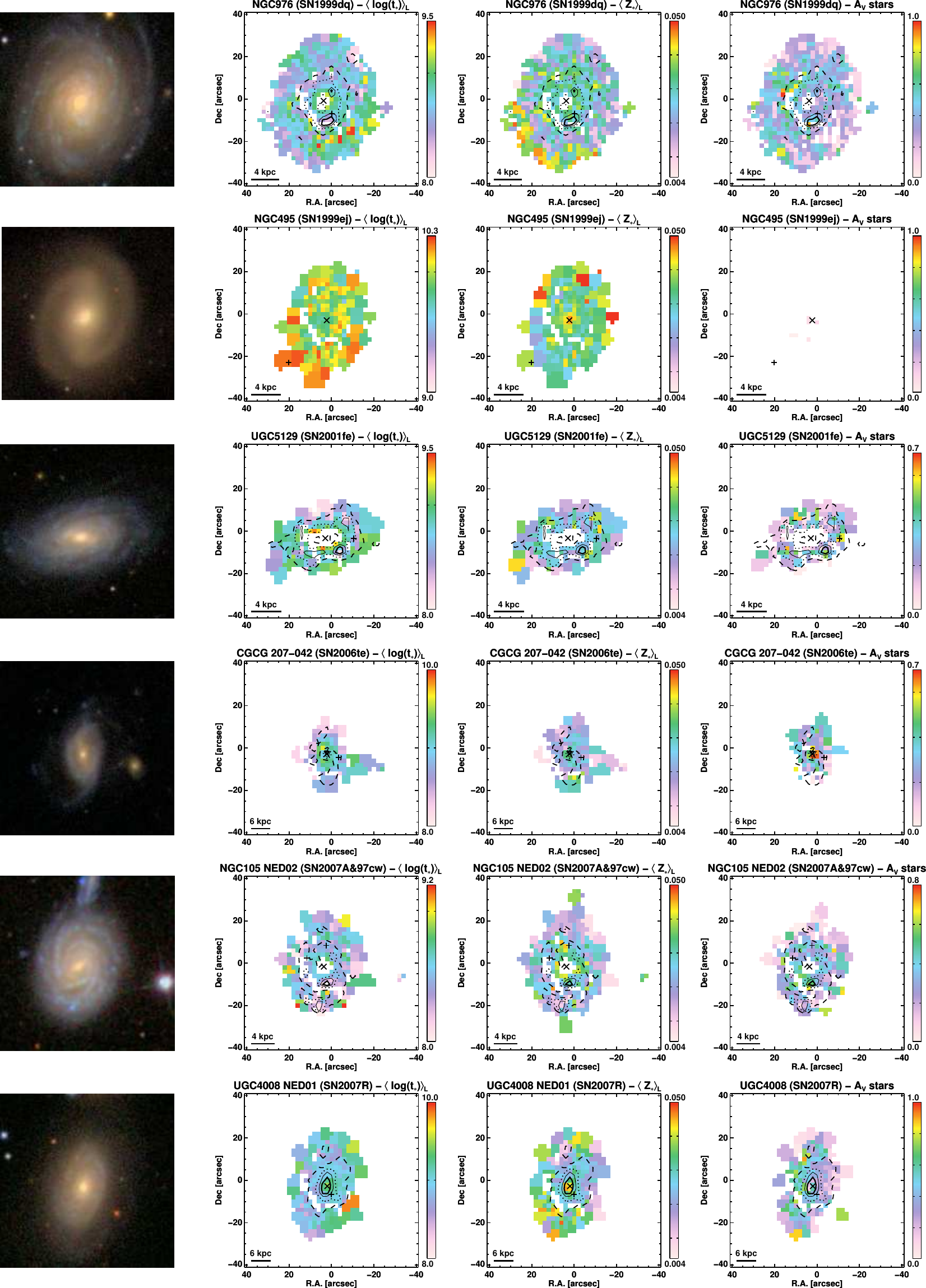}

\caption{From left to right: SDSS color images of the galaxies, the light-weighted average stellar population age $\langle \log t_\ast\rangle_{\rm L}$ and metallicity $\langle Z_\ast\rangle_{\rm L}$, and the 
visual extinction $A_{\rm V}$ determined by {\tt STARLIGHT} fits to the stellar spectra.}
\label{f:popaveL}
\end{figure*}
}

\subsubsection{Mean stellar age and metallicity}

\begin{table*}
\caption{Stellar population metallicity and age estimates from the {\tt STARLIGHT} fits. 
} 
\label{t:stm}
\begin{tabular}{@{}llccccccccc@{}}
\hline
\hline\noalign{\smallskip}
SN	 &   Host galaxy &  \multicolumn{4}{c}{total galaxy/minus AGN$^a$} &  &   \multicolumn{4}{c}{at the SN position$^b$}     \\
 \cline{3-6}\cline{8-11}\noalign{\smallskip}
 	    &      	        & $\left\langle\log(t_\ast)\right\rangle_M$  & $\left\langle Z_\ast\right\rangle_M$ &  $\left\langle\log(t_\ast)\right\rangle_L$  
	    & $\left\langle Z_\ast\right\rangle_L$   &   & $\left\langle\log(t_\ast)\right\rangle_M$  & $\left\langle Z_\ast\right\rangle_M$ & 
	     $\left\langle\log(t_\ast)\right\rangle_L$  & $\left\langle Z_\ast\right\rangle_L$  \\ 
\hline\noalign{\smallskip}
1999dq 	&  NGC 976    	 &   9.89/9.96 & 0.039/0.036   &   8.82/8.80 & 0.029/0.030  & &   9.75 & 0.042  &   8.82 & 0.030     \\ 
1999ej 	&   NGC 495      &  10.17      & 0.045         &   9.96      & 0.038	    & &  10.14 & 0.042  &  10.03 & 0.037     \\ 
2001fe 	&  UGC 5129      &   9.66/9.96 & 0.038/0.023   &   8.95/8.92 & 0.029/0.026  & &   9.45 & 0.037  &   8.71 & 0.031     \\ 
2006te 	&  CGCG 207-042  &   9.46      & 0.029         &   8.56      & 0.022	    & &   9.55 & 0.027  &   8.83 & 0.022   \\ 
2007A  	&  NGC 105 NED02 &   9.86/9.80 & 0.030/0.033   &   8.53/8.51 & 0.026/0.031  & &   9.78 & 0.040  &   8.48 & 0.029    \\ 
1997cw 	&  NGC 105 NED02 &   9.86/9.80 & 0.030/0.033   &   8.53/8.51 & 0.026/0.031  & &   9.77 & 0.039  &   8.43 & 0.028   \\ 
2007R  	&  UGC 4008 NED01&   9.68      & 0.042         &   9.01      & 0.031	    & &   9.96 & 0.044  &   9.11 & 0.030     \\ 
\hline 
\end{tabular}\\
\tablefoottext{a}{'total galaxy' - values derived by fitting the spectra formed by summing (un-weighted)
 \emph{all} spectra in the data cubes, 'minus AGN' - with the AGN affected spaxels excluded.}
\tablefoottext{b}{interpolated from the maps in Fig.\,\ref{f:popaveL}.}

\end{table*}

\onlfig{8}{
\begin{figure*}[t]
\centering

\includegraphics [width=18cm]{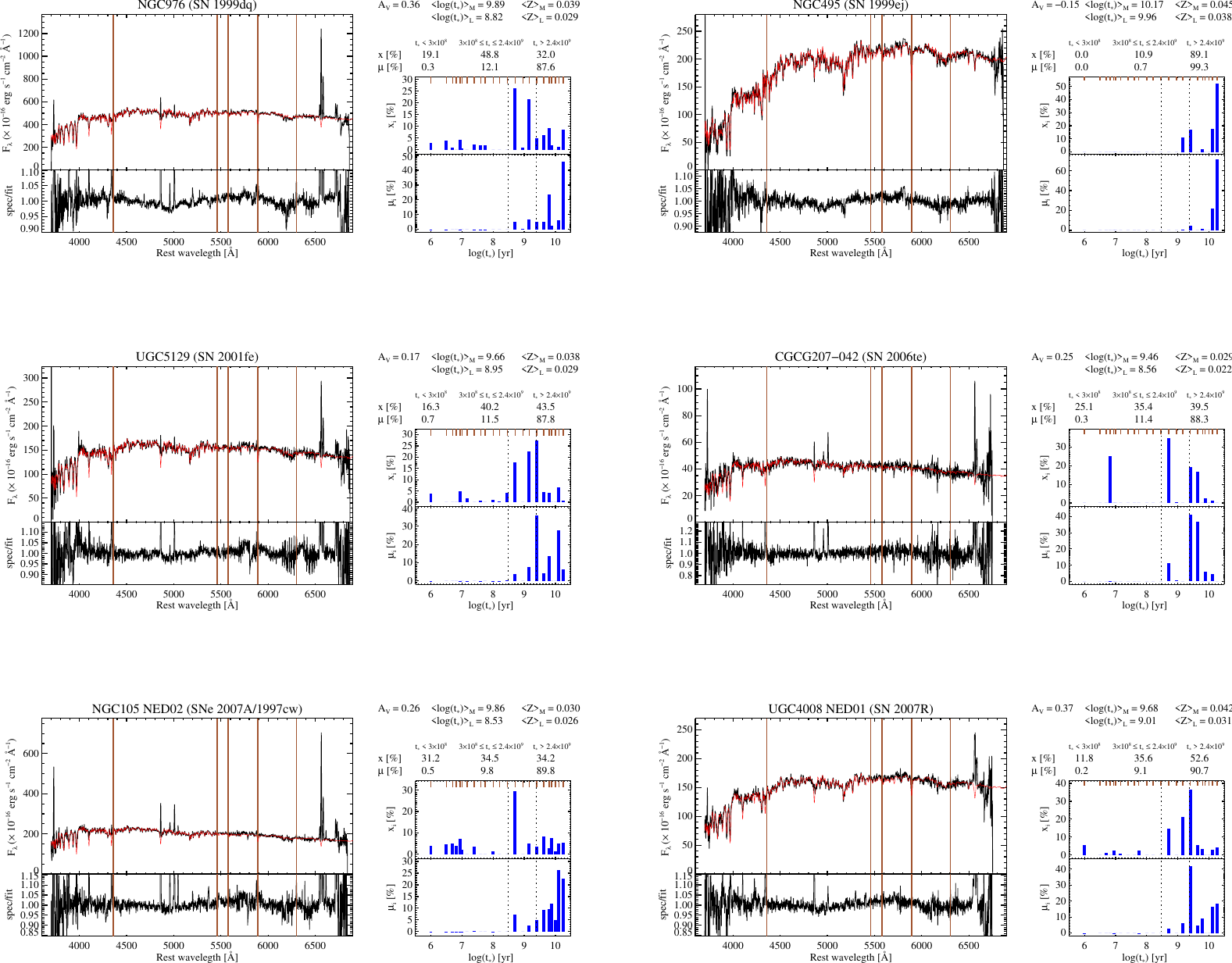}

\caption{{\tt STARLIGHT} fits to the total galaxy spectra. For each galaxy we also show the ratio between the observed and
the best-fit spectrum. The vertical brown lines in the spectrum panels show the location of the strongest night-sky lines, which could not be cleanly 
subtracted and the corresponding wavelength regions were excluded from the fits.
To the right of the plots are shown the population vectors and the mass-fraction vectors along with the 
extinction, the mass- and light-averaged age and metallicity, and the contribution of 
young (age $<$ 300 Myr), intermediate (300 Myr $<$ age $<$ 2.4 Gyr) 
and old (age $>$ 2.4 Gyr) stellar populations. The short brown bars show the ages of the SSPs used in the fits and the two vertical dotted lines separate the 
young, intermediate and old  populations. }
\label{f:totfit}
\end{figure*}
}

 Table\,\ref{t:stm} lists the mass- and light-weighted mean stellar population age and metallicity determined 
 from fitting the total galaxy spectrum formed as the sum of \emph{all} spaxels with (these fits are shown in Fig.~\ref{f:totfit}) and without the AGN-affected central spaxels.
The results show that all galaxies in our sample have a higher mean stellar metallicity than solar.  This is in accord with the findings from the emission lines analysis. The mean mass-weighted stellar age of the five emission line galaxies is $\sim5$ Gyr. 
\object{NGC~495}, which shows no emission lines, has an older stellar population of about 12 Gyr.
These values can be compared with studies based on total galaxy spectra, e.g. obtained with drift-scanning with a long-slit of local galaxies \citep[e.g.,][]{2005ApJ...634..210G} or spectroscopy of high-redshift galaxies when practically the whole galaxy light falls into the slit. 
Note that the residuals show a large-scale pattern with a full amplitude of up to $\sim$4\% (Fig.~\ref{f:totfit}). This signals either a problem with the relative flux calibration 
of the observed spectra or a problem in the SSP models. At present it is difficult to quantify what effect this would have on the results that are based on the spectral fitting.

\onlfig{9}{
\begin{figure*}[t]
\centering
\includegraphics [width=16cm]{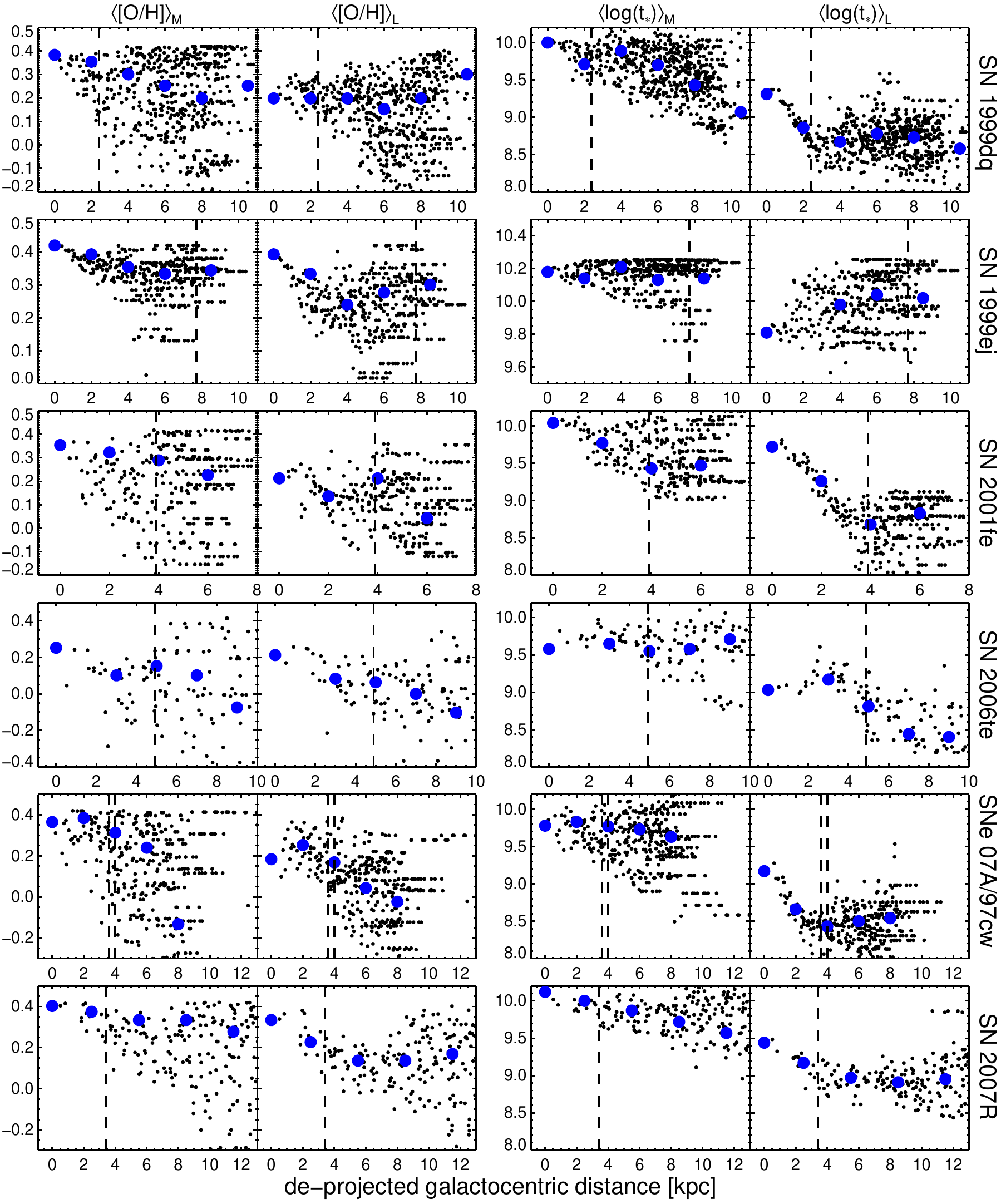} 

\caption{Mass- and light-weighted average stellar populations  metallicity (left panel) and age (right panel) from the {\tt STARLIGHT} fits
as function of the de-projected galactocentric distance.
The small black dots are the measurements on the individual spaxels and the large blue dots are from the azimuthally 
averaged spectra.}
\label{f:popaverad}
\end{figure*}
}

Figure~\ref{f:popaveL} shows the mass- and light-weighted mean stellar population age and metallicity maps of the six galaxies.
 In Table\,\ref{t:stm} are given the measurements for the total galaxy and at the SN position. 
The maps show considerable scatter and it is difficult to identify clear structures in them. Many spaxels that indicate high metallicity  
appear in the outer parts. This is most likely not real but rather a result of the insufficient S/N of the spectra even after applying the
 Voronoi binning. Nevertheless, there may be a slight increase of the metallicity toward the center, especially in 
\object{NGC~495} and \object{UGC~4008~NED01}. The same is also true for the stellar age maps, and again there 
is an indication of an older stellar population toward the nucleus, which can be expected. 

To investigate the matter in more detail, we plot in Fig.~\ref{f:popaverad} the the mass- and light-weighted mean stellar population age and metallicity measurements 
as a function of the de-projected galactocentric distance. Unfortunately, the plot confirms that the measurements from the individual spaxels spectra
show too large scatter. Unlike the ionized gas metallicity measurements, which show a small scatter of $\leq$0.05 dex at a given radius (Fig.~\ref{f:rad_met}),
the stellar metallicities estimated from the {\tt STARLIGHT} fits show scatter as large as 0.3 dex, for example at a radial distance of 6 kpc in 
 \object{NGC~976}, \object{UGC~5129}, and \object{NGC~105 NED02}. The age estimates also show considerable scatter.

\begin{table*}[!ht]
\caption{Compressed population vectors showing the contribution of the young (age $<$ 300 Myr), 
intermediate (300 Myr $<$ age $<$ 2.4 Gyr), and old (age $>$ 2.4 Gyr) stellar populations to the formation of the observed total galaxy spectrum. The total galaxy stellar masses  derived from our {\tt STARLIGHT} fits and
by \cite{2009ApJ...707.1449N} are also given.} 
\label{t:popfrac}
\begin{tabular}{@{}llcccrccccrccc@{}}
\hline
\hline\noalign{\smallskip}
SN	 &   Host galaxy &  \multicolumn{4}{c}{total spectrum}  &  &  \multicolumn{4}{c}{at the SN position$^b$} & & \multicolumn{2}{c}{log($M_{\star}$ [$M_{\sun}$])} \\
 \cline{3-6}\cline{8-11}\cline{13-14}\noalign{\smallskip}
&  &  Young & Inter. & Old & S/N$^a$ & &  Young & Inter. & Old & S/N$^c$ & & this work  & \cite{2009ApJ...707.1449N} \\ 
\hline\noalign{\smallskip}
1999dq 	&  NGC 976    	 & 0.19 & 0.49  & 0.32 & 81 &  & 0.19  & 0.41  & 0.40 & 84 & &  10.98 & 10.78 \\ 
1999ej 	&   NGC 495      & 0.00 & 0.11  & 0.89 & 54 &  & 0.00  & 0.00  & 1.00 & 33 & &  10.85 & $\dots$   \\ 
2001fe 	&  UGC 5129      & 0.16 & 0.40  & 0.44 & 97 &  & 0.29  & 0.33  & 0.38 & 81 & &  10.22 & 10.22 \\ 
2006te 	&  CGCG 207-042  & 0.25 & 0.35  & 0.40 & 35 &  & 0.21  & 0.18  & 0.62 & 42 & &  10.25 & 10.31 \\ 
2007A  	&  NGC 105 NED02 & 0.31 & 0.35  & 0.34 & 78 &  & 0.36  & 0.38  & 0.26 & 82 & &  10.61 & 10.87 \\ 
1997cw 	&  NGC 105 NED02 & 0.31 & 0.35  & 0.34 & 78 &  & 0.38  & 0.38  & 0.24 & 82 & &  10.61 & 10.87 \\ 
2007R  	&  UGC 4008 NED01& 0.12 & 0.36  & 0.53 & 54 &  & 0.14  & 0.29  & 0.57 & 85 & &  11.10 & 10.98 \\ 
\hline 
\end{tabular}\\
\tablefoottext{a}{S/N of the total galaxy spectra.}
\tablefoottext{b}{values at the SN radial distance interpolated from the radial dependencies derived from the azimuthally averaged spectra (Fig.~\ref{f:poprad}).
\tablefoottext{c}{S/N of the azimuthally averaged spectrum that is closest to the radial distance on the SN.}
}

\end{table*}

 The analysis of the emission lines shows that most of the ISM properties have 
a well-defined axial symmetry. One can expect this to be also the case for the stellar populations and 
hence asymmetries are unlikely  to be responsible for the observed scatter in the outer parts of the galaxies. 
 The scatter clearly increases with the radial distance (Fig.~\ref{f:popaverad}), suggesting that 
 the lower S/N of the spectra in the outer parts of the galaxies is causing it. 
 After the Voronoi binning the analyzed spectra a have minimum S/N$\sim$15-20 at 4600\AA. The large scatter that
 we observe in the derived quantities demonstrates the limitations of the full-spectrum fitting technique in the low-S/N regime and 
 suggests that an S/N significantly higher than 20 is needed to achieve reliable results.

\onlfig{10}{
\begin{figure*}[t]
\centering

\includegraphics [width=15.2cm]{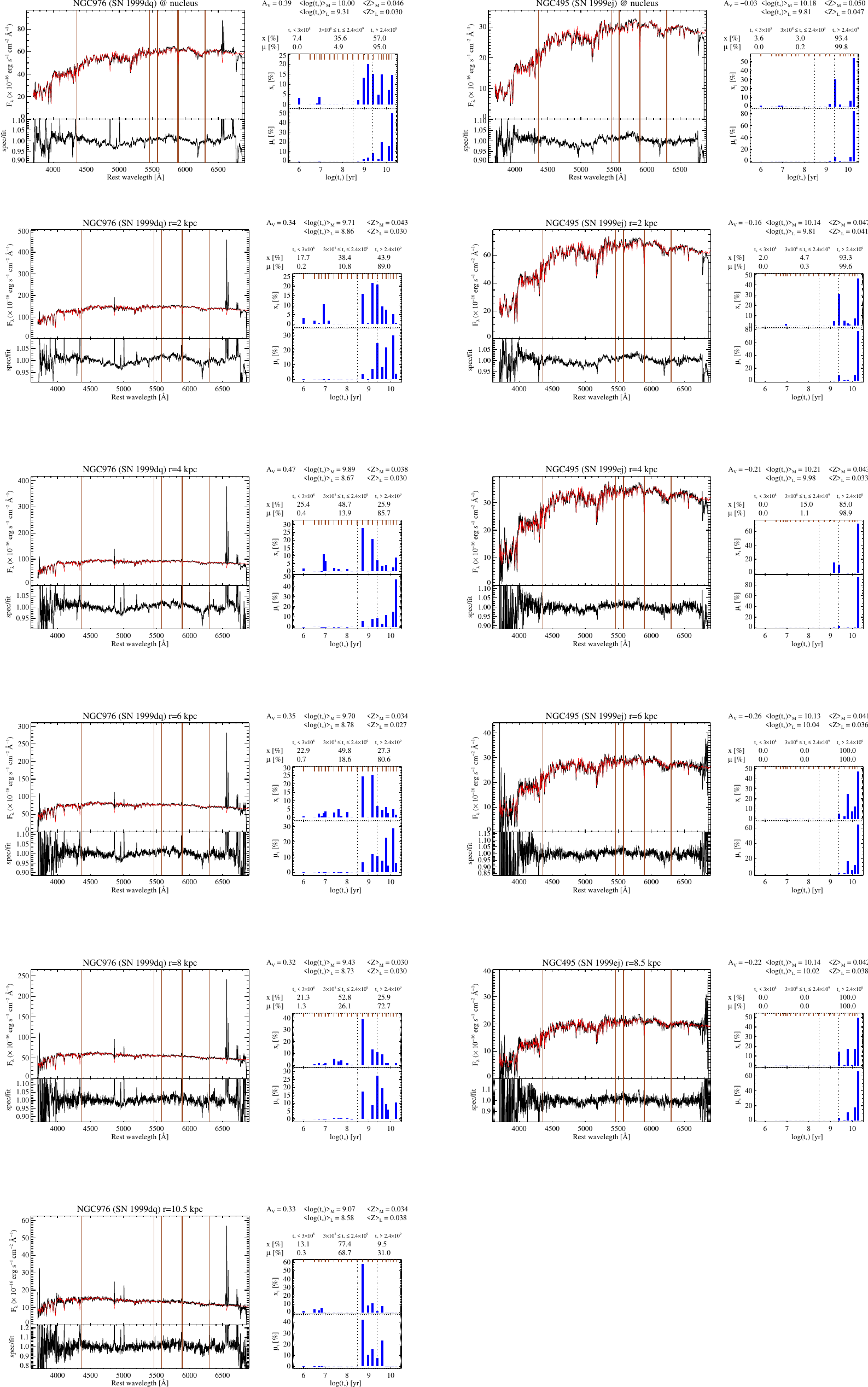}

\caption{Same as in Fig.~\ref{f:totfit}, but for the azimutally averaged spectra.}
\label{f:rfits1}
\end{figure*}
}

\onlfig{11}{
\begin{figure*}[t]
\centering

\includegraphics [width=17cm]{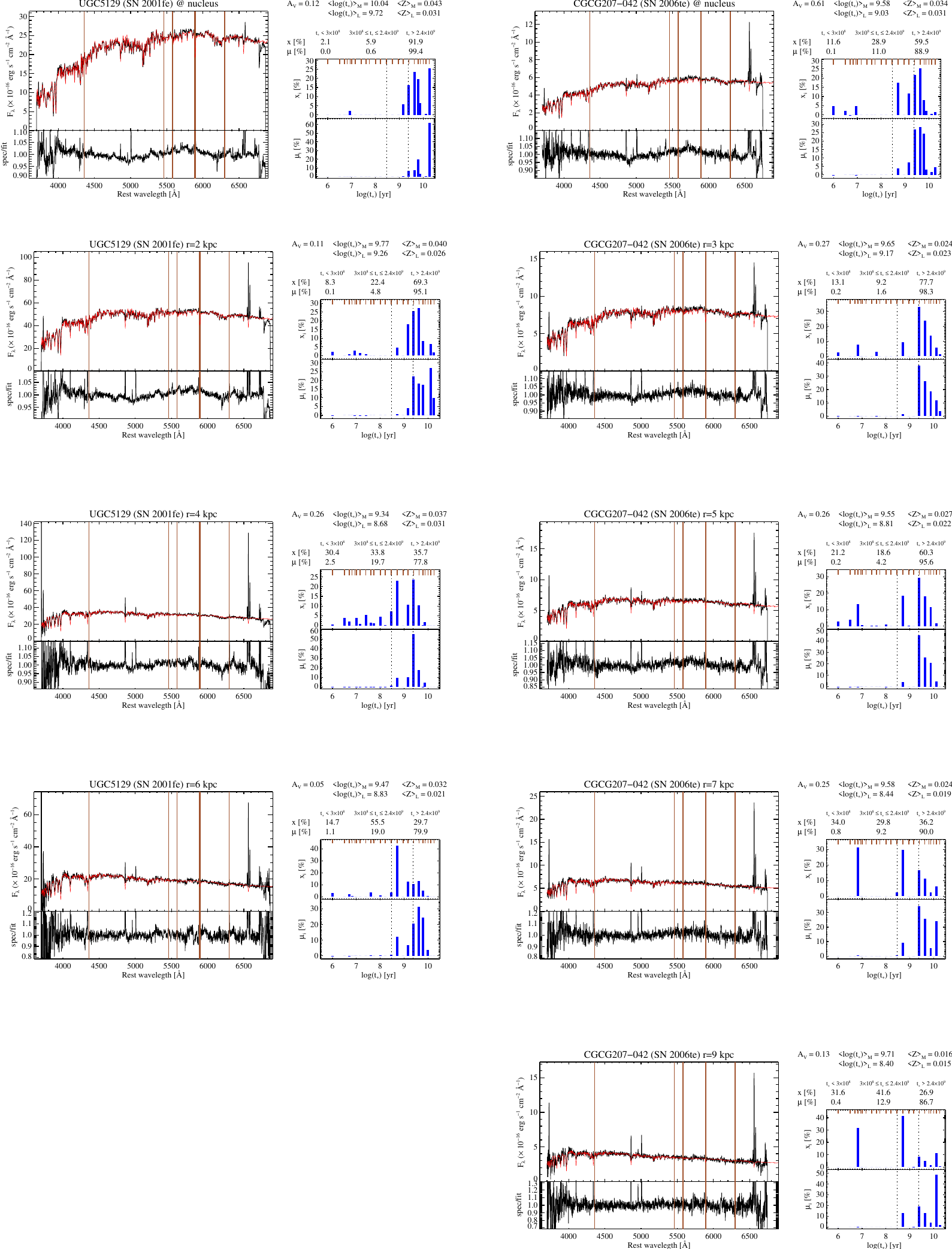}

\caption{Same as in Fig.~\ref{f:totfit}, but for the azimutally averaged spectra.}
\label{f:rfits2}
\end{figure*}
}

\onlfig{12}{
\begin{figure*}[t]

\includegraphics [width=17cm]{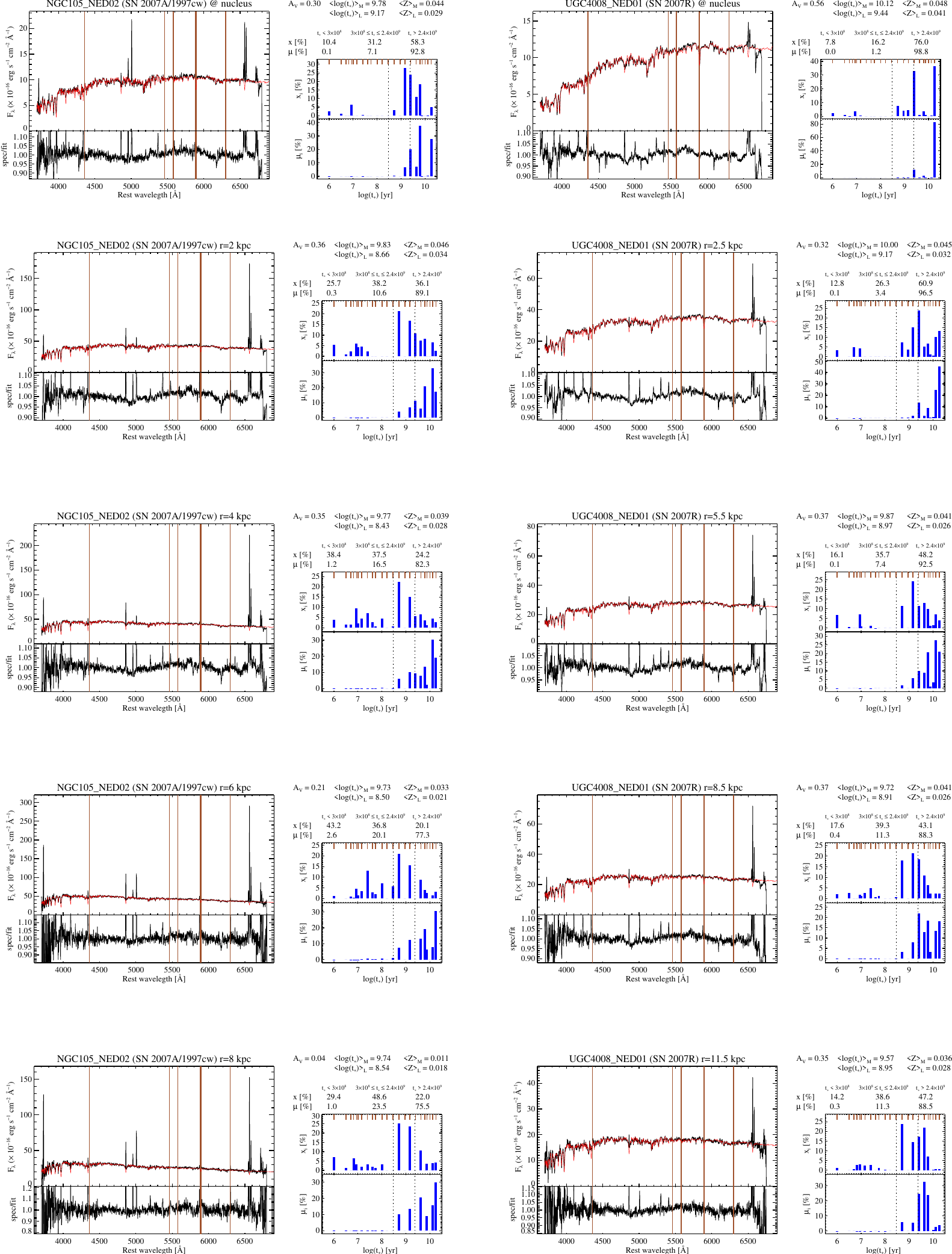}

\caption{Same as in Fig.~\ref{f:totfit}, but for the azimutally averaged spectra.}
\label{f:rfits3}
\end{figure*}
}

Given the large scatter of the  measurements from the individual spaxel spectra, interpolating at the location of the SNe from the 
2D maps is not recommended. An alternative approach is to use the measurements obtained from the azimuthally averaged spectra and interpolate them 
at the radial location of the SNe. This approach is better when there is evidence that the galaxy properties are symmetric around the nucleus.
In Fig.~\ref{f:popaverad} the blue symbols show the values estimated from the fits of the azimuthally averaged spectra and the vertical dashed lines show
the radial distance of the SNe. The corresponding fits are shown in Figs.~\ref{f:rfits1}-\ref{f:rfits3}. The mean age and metallicity show a smooth radial 
dependence. In some cases the  metallicity derived from the azimutally averaged spectra suggests negative gradients of up to $-0.03$ dex\,kpc$^{-1}$. However, given the large uncertainty with which the stellar metallicity is estimated ($\geq0.2$ dex), the significance of these gradients is difficult to assess. 
The mean ages qualitatively show
the same behavior with decreasing age outward. We note that the different types of weighting, mass or light, lead to 
different radial dependencies, with the light-weighted quantities showing stronger variation. The metallicity and the age 
at the locations of the SNe linearly interpolated from these radial dependencies are given in Table\,\ref{t:stm}.
 It is also worth mentioning that the light-weight quantities 
appear to have a slightly lower scatter, most pronounced in the inner regions where the spectra have a higher S/N. In most cases the light-weighted 
metallicies are lower than the mass-weight ones. The light-weighting gives much more weight to the younger stellar 
population and this result may imply that the younger populations have lower metallicity.

\onlfig{13}{
\begin{figure*}[!ht]
\centering

\includegraphics [width=17cm]{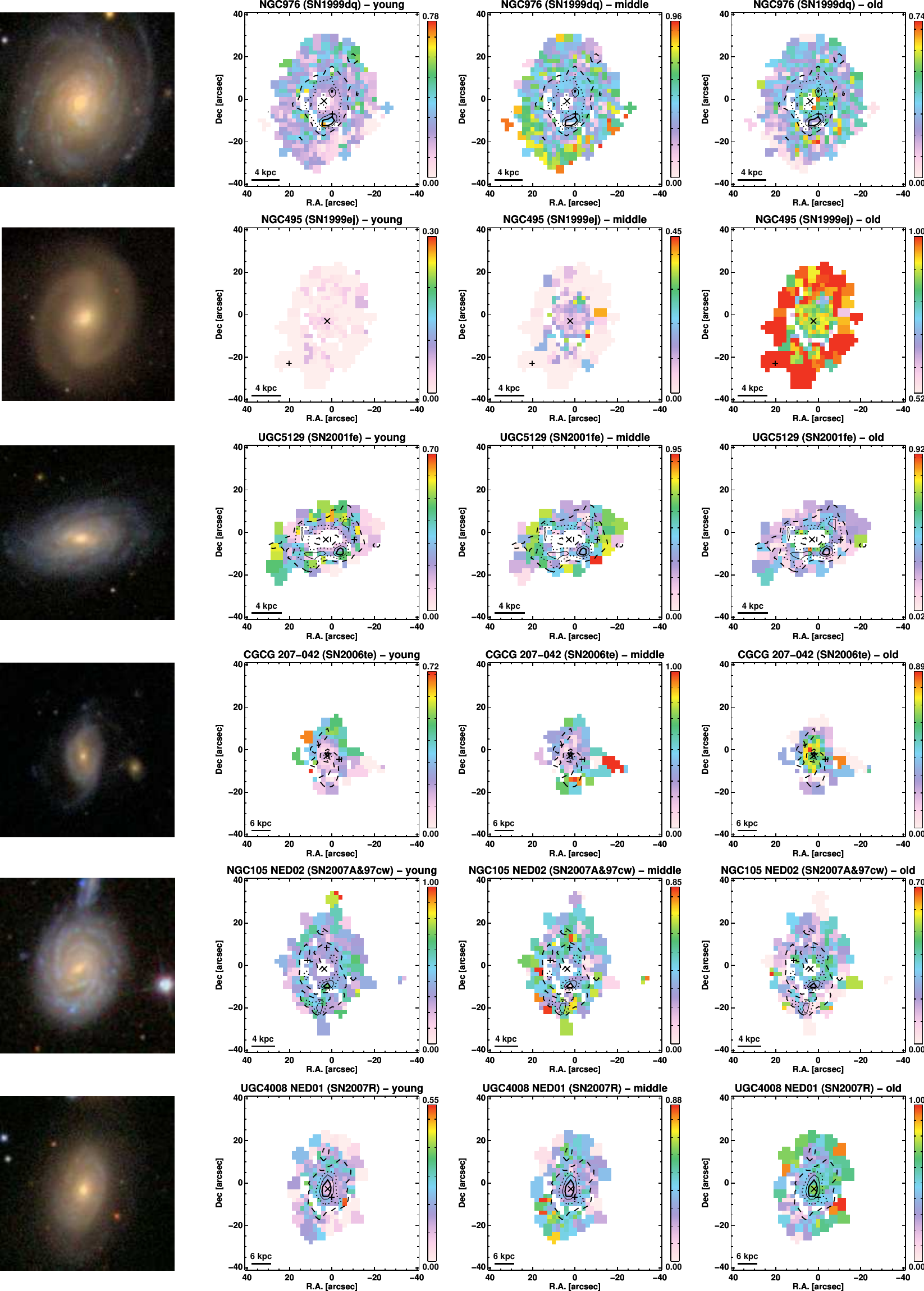}

\caption{From left to right: SDSS color images of the galaxies and the maps of the fraction of young (age $<$ 300 Myr), intermediate (300 Myr $<$ age $<$ 2.4 Gyr) and old (age $<$ 2.4 Gyr) stellar populations.}
\label{f:popfrac}
\end{figure*}
}

\subsubsection{Binned population vectors}

The 2D maps representing the fractional contribution of the young (age $<$ 300 Myr), intermediate (300 Myr $<$ age $<$ 2.4 Gyr), 
and old (age $>$ 2.4 Gyr) stellar populations are shown in Fig.\,\ref{f:popfrac}. In Fig.~\ref{f:poprad} 
the measurement from the individual spaxel spectra and the 
azimutally averaged spectra are plotted {\it vs.} the de-projected galactocentric distance. 
The measurements from the individual spectra again show 
considerable scatter in the outer parts of the galaxies. For this reason we again estimated the 
values at the radial distance SN from the azimutally averaged spectra and not from interpolation
 of the 2D maps. The estimated stellar population fractions at the SN radial distances are given
in Table\,\ref{t:popfrac} along with the values derived from the total galaxy spectra. The S/N of 
the spectra used to derive these values are also shown. The analysis shows that the five emission 
line galaxies contain stellar populations of different ages, including a considerable fraction of young stars.
In general, there is a clear trend of increasing the 
fraction of young stars with the radial distance. Depending on the galaxy, at a distance of 4-8 kpc the trend
 is reversed and the fraction of 
young stars starts to decrease. In four of the galaxies the fraction of old stellar populations
 monotonically increases toward the galaxy nucleus,
which is expected for most star-forming spiral galaxies. 
The exception is \object{CGCG~207-042}, the host of \object{SN~2006te}, which shows 
a decrease of the fraction of old stellar populations toward the center.
\object{NGC~495} is dominated by old stellar 
populations with possibly a small fraction of younger stars in the central few kpc.
 The contribution of the younger population is small, however, and its presence cannot be confidently confirmed.
 The behavior of the intermediate age  stellar populations is the opposite
to that of the old ones.

\cite{2005MNRAS.358..363C} showed that the compressed population vectors can be recovered with an accuracy better 
than $\sim$10\% for S/N$>$10. However,
considering the uncertainties involved in the computation of the SSP models as well as other uncertainties such 
as the correlations between the fitted parameters, the relative flux calibration and the dust extinction laws in 
the galaxies, the accuracy is probably no better than $\sim$10\%. 
This is also supported by the level of the scatter in Fig.~\ref{f:poprad}. 
In this context, the population vectors at the locations of SNe  \object{1999dq}, \object{2007A}, \object{1997cw}, and \object{2007R} 
are un-distinguishable from those of the whole galaxies (Table\,\ref{t:popfrac}). At the location of \object{SN~2006te}  
there is a larger contribution from old populations at the expense of the intermediate age, while the fraction of young stars is the same as for the whole galaxy.
For \object{SN~2001fe} there is marginal evidence for an increased contribution of a young population at the position of the SN.
The host of \object{SN~1999ej} formed the bulk of its stars about 13 Gyr ago followed by a less intense star-forming period about 2 Gyr ago.
At the distance of  \object{SN~1999ej} we only find evidence for the older population.

\onlfig{14}{
\begin{figure*}[t]
\sidecaption
\includegraphics [width=12cm]{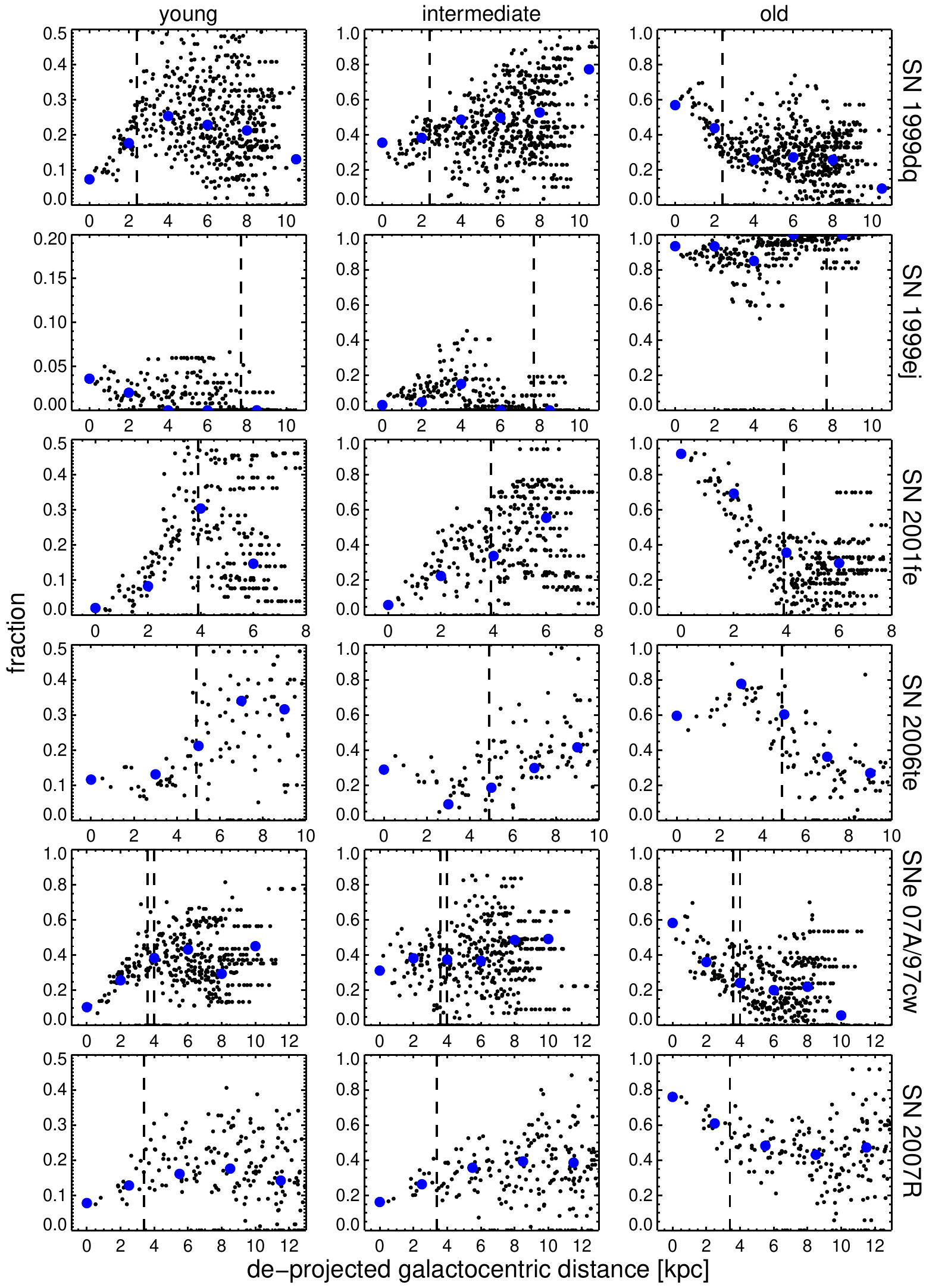} 
\caption{Compressed population vectors corresponding to the contribution of young (age $<$ 300 Myr), 
intermediate (300 Myr $<$ age $<$ 2.4 Gyr), and old (age $>$ 2.4 Gyr) stellar population
to the formation of the observed spectra as function of the de-projected galactocentric distance. 
The small black dots are the measurements obtained from the individual spaxels and the large blue dots are from the azimuthally 
averaged spectra.}
\label{f:poprad}
\end{figure*}
}

\subsubsection{Stellar kinematics}

The velocity dispersion maps derived from the {\tt STARLIGHT} fits show a simple morphology with a
single peak centered at the galaxy nucleus.  In Fig.~\ref{f:velmaps}
the velocity fields of the stars in the five emission line galaxies are compared to the velocity fields 
derived from the H$\alpha$ emission line. The two maps are very similar and small systematic 
differences are only revealed  after subtracting the two maps (the last column in Fig.~\ref{f:velmaps}).
Evidently, the gas rotates faster  in the central regions than the stars,  with the difference being largest 
in \object{UGC~4008~NED01}. These differences between the rotation of stars and ionizied gas in the central regions of galaxies are well-known and 
have been extensively studied \citep[see, e.g., ][and references therein]{2004A&A...424..447P}.
We note that none of the galaxies shows a sign of counter-rotating gaseous disk
\citep[e.g.,][]{1996ApJ...458L..67B,1992ApJ...394L...9R}.

The stellar velocity fields were also analyzed with the methods of \cite{2006MNRAS.366..787K}.
As with the H$\alpha$ velocity map, within the uncertainty we also found no evidence 
for deviations from pure disk rotation.

\subsubsection{Current stellar mass}

An estimate of the present-day stellar mass of the galaxies was obtained from the {\tt STARLIGHT} fits of the total galaxy 
spectra. The fits are shown in Fig.~\ref{f:totfit} and the masses are given in Table\,\ref{t:popfrac}. All the galaxies have masses exceeding $2\times10^{10}\,M_{\sun}$ 
and can be classified as quite massive.
The fairly high metallicity that we derived for both the 
ionized gas and the stellar component are therefore in line with the expectation from the mass-metallicity relation, e.g. \cite{2004ApJ...613..898T}.
We note that the values that we obtain are very close to those of \cite{2009ApJ...707.1449N}, which were obtained by a different methodology
(see Sec.~\ref{sec:ha}).

\onlfig{15}{
\begin{figure*}[t]
\centering

\includegraphics [width=18cm]{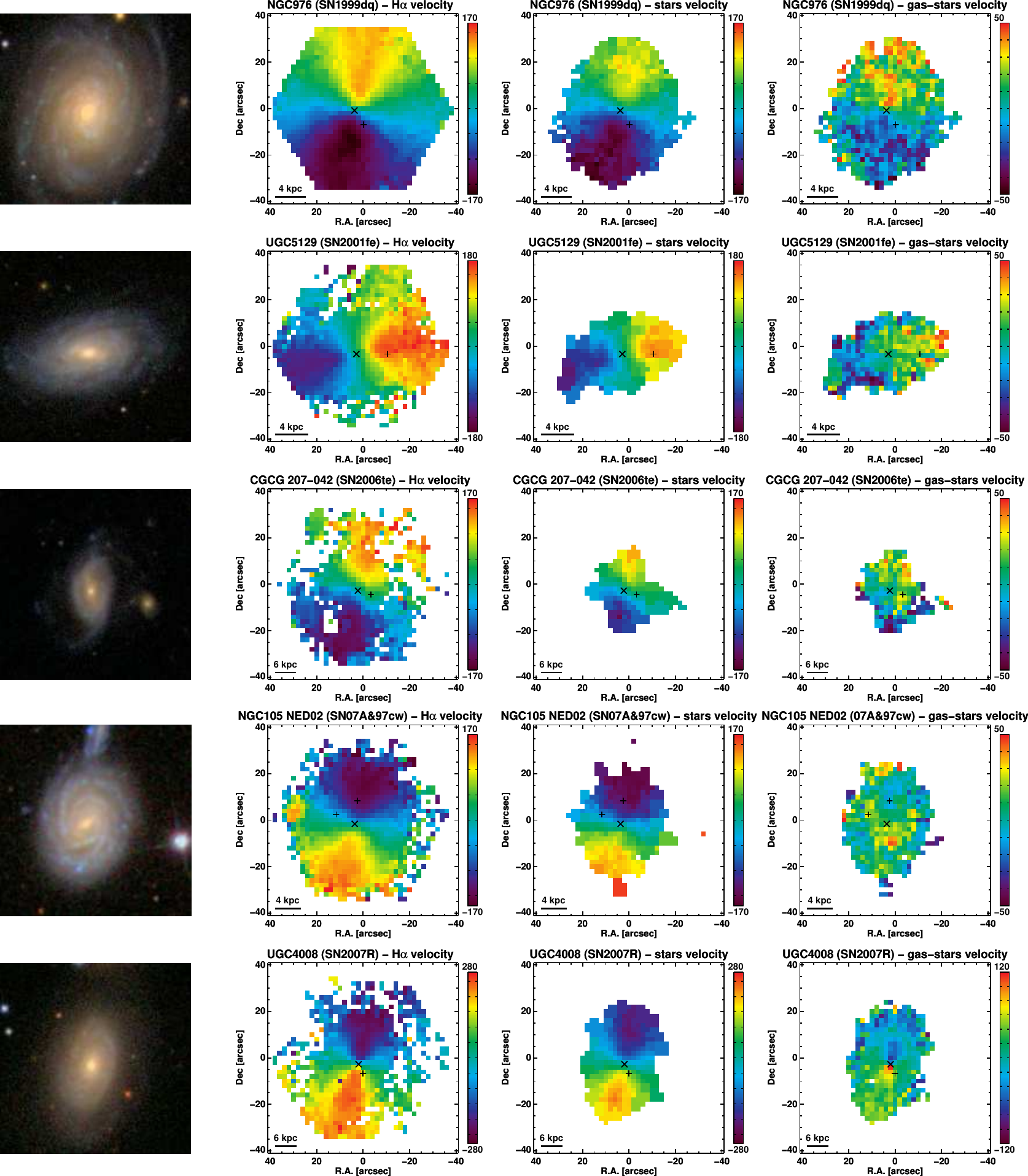}

\caption{From left to right for each galaxy we show the color SDSS image,  H$\alpha$ velocity map, the star velocity map, and the 
difference between them. The $x,y$ coordinates are in arcsec with respect to the map centers. The orientation of the images is north -- up, east -- left.}
\label{f:velmaps}
\end{figure*}
}

\section{Discussion}

\subsection{Galaxy mass and metallicity}

We used IFU spectroscopy to derive the spatially resolved properties of six face-on spiral galaxies that hosted seven nearby SNe~Ia. 
The masses of the galaxies derived from the analysis of the total spectra with the {\tt STARLIGHT} code are all higher than $2\times10^{10}\,M_{\sun}$.
Recently,  \citet{2010ApJ...715..743K}, \citet{2010MNRAS.406..782S},  and  \cite{2010ApJ...722..566L} have claimed 
that the residuals from the best-fit Hubble line correlate with the SN host stellar mass. Furthermore,  \citet{2010MNRAS.406..782S} proposed to 
incorporate into the cosmological SN~Ia analyses two different absolute peak magnitudes for SNe in hosts with masses lower or higher than $10^{10}\,M_{\sun}$; after the "lightcurve width -- luminosity" and color corrections the SNe in the more massive hosts are found to be $\sim0.06-0.09$ mag brighter than their counterparts in lower mass hosts.
The galaxies in our sample fall into the high-mass/low-specific SFR bins  defined by  \citet{2010MNRAS.406..782S}. Accordingly, one can expect the SN
in these galaxies to have on average negative Hubble residuals. From Table~\ref{t:snprop} one can see  that four of the 
SNe have significant positive residuals ($>2\sigma$). The other three have negative residuals, but only one of them 
is bigger than the uncertainly. The mean weighted residual is positive, $+0.07\pm0.22$; however, one should keep in mind that we used only very few  SNe in our analysis.

The cause of the apparent dependence of the SN~Ia luminosity on the host galaxy stellar mass is still unclear. Theoretical investigations have 
shown that various parameters of the exploding WD, such as its metallicity, C/O ratio, central density, and progenitor age can affect the
amount of $^{56}$Ni synthesized in the explosion to a different degree and hence the SN luminosity
 \citep[see, e.g.,][ and references therein]{2003ApJ...590L..83T,2006A&A...453..203R,2009ApJ...691..661H,2010ApJ...711L..66B}. 
Among these parameters, the metallicity is known to correlate with the 
galaxy mass \citep[see, e.g., ][]{2004ApJ...613..898T} and is likely to have the strongest impact. 
Our analysis of the emission line fluxes and the stellar populations revealed that the galaxies in our sample have on
 average solar and higher metallicity (Tables~\ref{t:oh} and \ref{t:stm}). This is not surprising because the galaxies are
quite massive and by the virtue  of the mass-metallicity relation \citep[see, e.g., ][]{2004ApJ...613..898T} may be expected to have high metallicities. For five of the SNe, the ISM metallicity measured at the location of 
  the SN is higher than the galaxy average by about $\sim0.1$ dex (Table~\ref{t:oh}). 
  This can be explained by the presence of
radial metallicity gradients and our target selection criteria.
Figures~\ref{f:rad_met} and \ref{f:popaverad} show that the galaxies in our sample 
have radial metallicity gradients. At the same time, the selection criterion that the 
SNe are located on a high surface brightness location in the galaxies led to a 
SN sample that is biased toward SNe close to the galaxy nuclei. Together with the presence of the metallicity gradients, this resulted in most of the SNe 
being at locations with higher-than-average metallicity within the galaxies \citep[see also][]{2005PASP..117..227K}.

\begin{figure}[!t]
\centering
\includegraphics*[width=8cm]{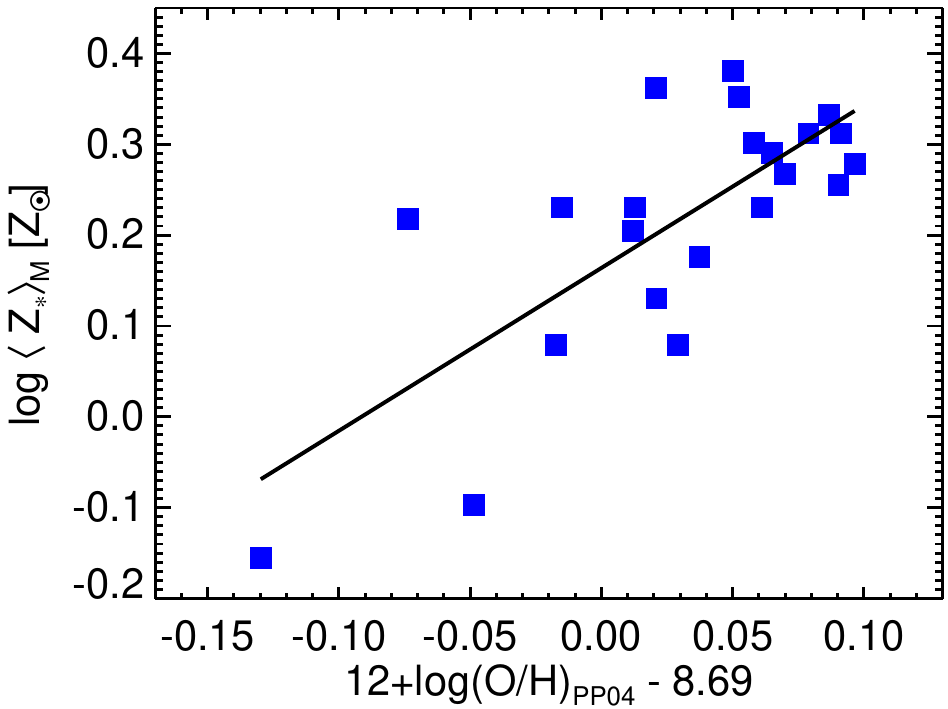} 
\caption{Mass-weighted stellar metallicity  {\it vs.}  gas-phase oxygen abundance estimated from the
azimutally averaged spectra.}
\label{f:met2}
\end{figure}

While the gas-phase metallicity is easier to estimate, a more relevant
 quantity is the stellar metallicity. Figure\,\ref{f:met2} shows the mass-weighted stellar metallicity vs. the gas-phase oxygen  abundance estimated from the azimutally averaged spectra. The two quantities appear to be correlated. The slope of the linear fit is $\sim1.8$ with a dispersion of $\sim0.1$ dex. Note that \cite{2005MNRAS.358..363C} also found that the gas-phase and the stellar metallicities are correlated from an analysis of a large sample of SDSS galaxies. However, the two relations are difficult to compare because the \cite{2005MNRAS.358..363C} analysis also included low-metallicity galaxies and galaxies in a somewhat higher redshift interval.

\subsection{The impact of the metallicity gradient}

\begin{figure}[!t]
\centering
\includegraphics*[width=8cm]{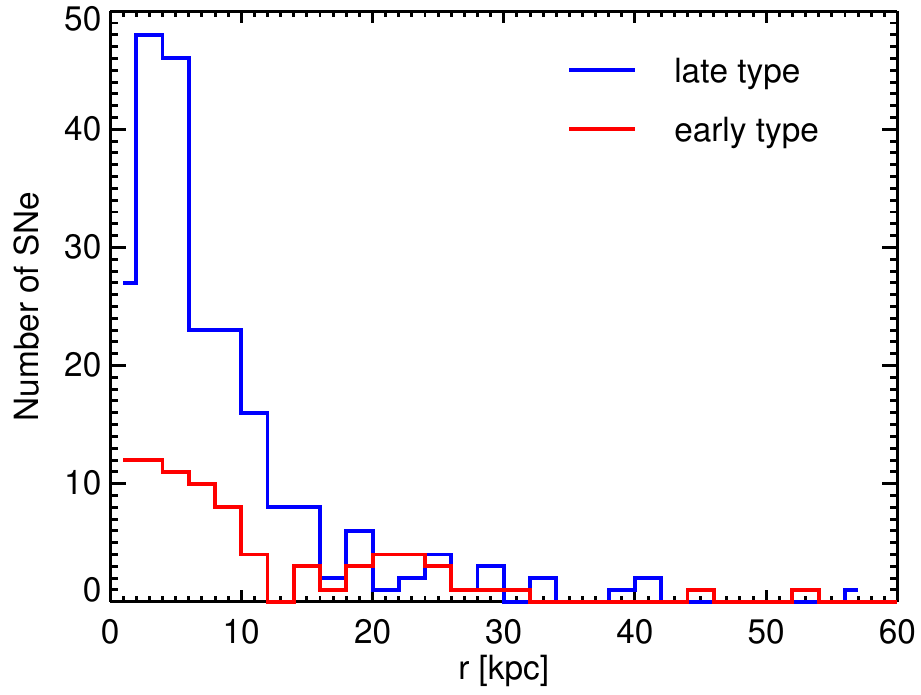} 
\caption{Distribution of the \emph{observed} galactocentric distance for a sample of 305 nearby SNe~Ia in late- and early-type hosts. 
Because the distances have not been de-projected these are the \emph{minimum} galactocentric distances. }
\label{f:rad_stat}
\end{figure}

The presence of abundance gradients in both spiral and elliptical galaxies is now a well-established fact
\citep[e.g.,][]{1994ApJ...420...87Z,1999PASP..111..919H}. If not taken into account, the gradients will affect any attempt 
to study the properties of SNe~Ia and/or their progenitors as a function of their host galaxy metallicity. The values of the gradients 
seen in the galaxies in our sample suggest that SN~Ia progenitors that form at radial distances greater than $\sim15$ kpc may have metallicities 
that are lower by a factor at least 2-3 than progenitors in the central parts. Studies of the radial distribution of SNe~Ia within their hosts galaxies
have shown that more SNe  explode in the central regions \citep[e.g.,][]{2000ApJ...542..588I,1997ApJ...483L..29W,1997AJ....113..197V,2007HiA....14..316B}.
However, SNe~Ia are also found at large  galactocentric distances in both spiral and elliptical galaxies. In Fig.~\ref{f:rad_stat} we show the 
distribution of the projected galactocentric distances  (PGD) for a sample of 305 SNe with modern CCD observations (observed after 1990) and with known
host galaxy type, redshift, and offset from the center. About 7\% of the SNe in spiral galaxies and 20\% in the ellipticals are found at PGD$>20$ kpc. 
Since the real galactocentric distances are always greater than, or equal to, the PGD, the above-mentioned fractions are lower limits. Therefore, a significant fraction of 
SNe may have progenitors with a metallicity that is much lower than that of the host average.

An important question is whether the present-day galaxy metallicity is a good proxy of the metallicity of SN~Ia progenitors. This
was recently studied by \cite{2011MNRAS.414.1592B}, who used simplified one-zone galaxy evolution models coupled with the
SN delay-time distribution (DTD) functions of \cite{pritchet08s} and \cite{2010ApJ...722.1879M}. The authors concluded that 
 the galaxy ISM metallicity is a good proxy for the SN progenitor metallicity and derived simple linear relations to
 estimate the progenitor metallicity from the present-day host metallicity.  However, \cite{2011MNRAS.414.1592B} did not include 
 metallicity gradient, and more importantly, its possible evolution with time. There is growing evidence that the disks in 
 late-type galaxies formed and evolved slowly under the constant inflow of metal-pool gas from the galactic halo.
The galaxy chemical evolution models and hydrodynamical simulations have shown 
 that the metallicity gradient evolves considerably during the last 10 Gyr of the galaxy evolution
\citep{2012arXiv1201.6359P,2009ApJ...696..668F,2006MNRAS.366..899N,2009MNRAS.398..591S,
2005MNRAS.358..521M,2001ApJ...554.1044C,1997ApJ...475..519M}. 
Although the exact results depend of the particular code and model used \citep{2012arXiv1201.6359P}, all
studies but one \citep{2001ApJ...554.1044C} show that the metallicity gradient was steeper in the past
and gradually flattens out to reach present-day values similar to those observed in local spiral galaxies. 
Recently, there has also been observational support for this conclusion. \cite{2011ApJ...732L..14Y} and \cite{2010ApJ...725L.176J}
reported metallicy gradients of $-$0.16\,dex\,kpc$^{-1}$ and $-$0.27\,dex\,kpc$^{-1}$ for galaxies at redshifts z=1.5 and z=2.0,
respectively. We note that the galaxy chemical evolution studies show 
that the mean disk metallicity has increased slowly by $\sim0.3-0.5$ dex during the last several Gyr. 
The gradients seen in the galaxies in our sample and in other galaxies at low and high redshift 
imply that the metallicity differences within the same galaxy  may exceed the cosmological 
increase of the mean metallicity. In addition, some studies have pointed out that the metallicity gradient in
the outermost parts of the galaxies may be steeper than in the inner disk \citep[see, e.g.,][]{2009ApJ...696..668F}.

The above studies highlight the complexity of estimating the metallicity of the SN~Ia progenitors from their
host galaxy present-day metallicity. The difference between the present host metallicity and the SN progenitor metallicity 
is a complex function of several factors, some of which are poorly understood and not very well constrained with observations: 
the radial distance at which the progenitor formed, the age of the progenitor, and the evolution of the metallicity gradient. For example, a progenitor that formed at large radial distance 
will have increasingly larger difference from the preset-day metallicity at the same radius as the 
progenitor ages.
Another uncertainty can be added if the galaxies have experienced major mergers and radial star migrations, which tend to flatted the metallicity gradient
 \citep[see, e.g.,][]{2010ApJ...721L..48K,2009MNRAS.398..591S}. This complexity may be the reason why the attempts to
correlate the Hubble residual with the host global metallicity have not led to conclusive results 
\citep{2008ApJ...685..752G,2009ApJ...691..661H,2005ApJ...634..210G,2011arXiv1110.5517D}.
Note however that \cite{2009ApJ...691..661H} did not directly measure the 
metallicity but rather estimated it from a mass-metallicity relation.

All SNe in our sample but one are within 5 kpc from the galaxy centers.
Generally, the chemical evolution models show
that the metallicity close to the galaxy nuclei changes least. Therefore,  the 
metallicity of the SN progenitors that formed near the center should be closer to the present-day 
galaxy metallicity compared to the progenitors that formed in the outer parts. 
 Together with the fact that we measured fairly high present-day metallicity at the locations of all SNe, 
this suggests that their progenitors did not form in metal-poor environments, unless they came from very old stellar
population with a long delay time.

\subsection{Star formation history}

Much of the recent progress on the question of SN~Ia progenitors has been achieved through studies of the SN rates.
It is now well-established that the SN~Ia rate depends on both the total stellar mass and the recent SFR 
in the host galaxy  \citep[e.g.,][]{2005A&A...433..807M,2005ApJ...629L..85S,2006ApJ...648..868S,2010ApJ...722.1879M,2010AJ....140..804B},
which led to a two component model for the SN~Ia rate, the co-called A+B model.
Along with the fact that SNe~Ia are also observed in old, passive galaxies, this points to the existence of at 
least two evolution channels for SNe~Ia associated with young and old stellar populations. 

Except for 
\object{NGC~495}, all other galaxies in our sample contain a considerable fraction of young stars
and strong H$\alpha$ emission, indicating ongoing star formation activity. 
The {\tt STARLIGHT} fits of the total galaxy spectra are shown in Fig.~\ref{f:totfit}. Except for \object{NGC~495}, all other galaxies show a similar 
pattern, namely, the population vectors $x_j$ show the largest contribution from SSPs with ages 0.5-5 Gyr.
Young populations, $\sim$50 Myr, are also confidently detected in all cases. From Figs.~\ref{f:popfrac} and \ref{f:popaverad} it can be seen that 
the fraction of young stars increases with increasing the radial distance.
It is known that the full-spectrum fitting techniques tend to estimate suspiciously large components with ages $\sim1$ Gyr  \citep[see, e.g.,][]{2007MNRAS.381..263A,2006MNRAS.365..385M,2007MNRAS.378.1550P} and one may ask whether the large contribution of SSPs of similar ages that we see in our analysis is real. \cite{2007MNRAS.381..263A} and \cite{2009RMxAC..35..127C} report that the problem disappeared once they switched from the original \cite{2003MNRAS.344.1000B} fitting basis based on STELIB to a new basis that uses the MILES spectral library. Because we also used the newer Bruzual \& Charlot basis based on MILES, our results are also likely unaffected by the above-mentioned problem.

We estimated the current SFR rate from the H$\alpha$ emission line flux (Table~\ref{t:sfr}). 
Because most of the ionizing photons are produced by massive, short-lived stars, the H$\alpha$ flux
is a tracer of the very recent star formation, $\leq 20$\,Myr. Another estimate of the SFR can be obtained from the {\tt STARLIGHT} fits
following the methodology described in \cite{2007MNRAS.381..263A}. We estimated the mean SFRs during the last 0.5 Gyr and last 50 Myr.
The values are given in Table~\ref{t:sfr}. The mean SFRs over the last 50 Myr are very similar to the estimates obtained from the H$\alpha$ flux; 
\cite{2007MNRAS.381..263A} have already demonstrated that there is a tight correlation between these two estimations using a large sample of SDSS galaxies. On the other hand, the mean SFRs during the last 0.5 Gyr are by a factor 3-5 higher than the SFR estimates from H$\alpha$ flux, but 
are similar to those of \cite{2009ApJ...707.1449N}. The only 
exception is  \object{UGC 4008 NED01}, for which we obtain a much higher value; note, however, that the confidence interval quoted by 
\cite{2009ApJ...707.1449N} has an upper limit higher by an order of magnitude than our estimate.  Note also that the model SEDs 
that \cite{2009ApJ...707.1449N} fitted to the broad-band photometry are based on eight galaxy models with pre-defined
SFHs, which were meant to represent the Hubble galaxy types plus one star-burst galaxy model. {\tt STARLIGHT} does not
assume any pre-defined SFH and the contributions of all SSPs are free parameters. Therefore, {\tt STARLIGHT} is much more flexible to
describe galaxies with arbitrary SFHs.

If the two components of the A+B model represent the contribution of two different channels to produce SNe~Ia, 
 we can estimate the probability from which channel the SNe in our galaxies were produced. With the A and B constants 
estimated by \cite{2006ApJ...648..868S} and  our measurement of
the galaxies' total mass and SFR,  the SNe in the five star-forming galaxies have an about equal chance to have come
from the young or the old channel.

\subsection{Correlation between the SN and host galaxy properties}

Despite the low statistics of our sample, we can still test correlations between the SALT2 $x_1$ parameter and the Hubble residuals  of the SNe with the various parameters that we
derived for the total galaxy and at the SN locations (Tables~\ref{t:sfr}-\ref{t:popfrac}).
We found no
statistically significant correlations. The small number of objects is certainly not ideal for this analysis, but the metallicity and
the masses of the galaxies in our sample also span quite a small range. To search for correlations it is necessary to expand the sample 
toward lower masses and metallicities.  This may not be an easy task because SNe~Ia in metal-poor galaxies in the local Universe are 
rare. Besides, the low-mass, low-metallicity galaxies tend to be faint and are more difficult to observe with a sufficient S/N.

In addition to the above concerns, when correlating the SN properties with the properties of their host galaxies at the location
of the SN one should always bear in mind that the SN progenitors may be very old stars \citep[e.g.,][]{2011MNRAS.412.1508M}.
The progenitor system may have migrated from its 
birth place and the galaxy properties at its present location may be different from those where the progenitor has formed.
\cite{2011MNRAS.412.1508M} discussed that random stellar motions will affect the SN progenitor and its surroundings in the same way.
As a result, the population that gave birth to the SN progenitor will also be present in the new SN location. 
Following the same argument, the radial star migrations should not have a significant effect either. 
Additionally, \cite{2009MNRAS.398..591S} found that within the central 5-10 kpc the radial star migration during the whole 
galaxy evolution is fairly small $\sim1.5$ kpc and increases to only $\sim3.5$ kpc in the outer parts.  
 
Another problem is related to the projection effects. Because the SN~Ia progenitors have a broad range of ages between $\sim100$~Myr and  10~Gyr, SNe can 
explode anywhere along the line of sight. An SN produced by an young progenitor would have most likely exploded in the galactic disk, where 
the newly formed stars and the ionized gas typically reside. In this case the progenitor star's metallicity should be close to that of
the ionized gas at the (projected) location of the SN. In the case of old progenitors, however, 
the SN may have exploded in the galactic halo and the metallicity of its progenitor may be very different from the gas. 
Similarly, the stellar continuum at a given spaxel is the sum of all star light along the light of sight. 
All these effects indicate that the correlation of the SN properties with the properties of its local environment 
is not unambiguous.

Clearly, to study the correlation between the SN properties and its local environment, a large, 
unbiased sample of galaxies observed with a large-field IFU spectrograph is needed. Our sample 
of seven SNe~Ia in six galaxies is not large enough to draw any conclusion. The ongoing 
CALIFA\footnote{\url{http://www.caha.es/CALIFA/public_html/}} survey \citep{2011arXiv1111.0962S} will provide IFU 
observations of about 600 galaxies at redshift $z\sim0.02$. CALIFA uses the same instrument as the observations presented in
this paper with similar setup and exposure times, and will provide data of similar quality as ours. The CALIFA targets  
are selected from a larger pool of about 1000 galaxies based only on the visibility of the targets at the time of the observations. 
Many of these galaxies are known to have hosted SNe. In addition, there are 
several ongoing large-field SN searches with that will potentially discover many new SNe in CALIFA-targeted galaxies.
Thus, the CALIFA survey will provide a solid base to further expand the studies SNe~Ia properties as a function of their local environment
to all SN types. In addition, CALIFA will provide the full galaxy spectra that will allow one to avoid the 
aperture  effects to which the SDSS spectroscopy is subject. At low redshift the 3\arcsec-diameter fibers of the SSDS spectrograph
cover only the galaxy nucleus, whose properties may be very different from those of the disk and may not be representative for the 
environment of the SNe that exploded elsewhere  in the galaxy, for example because of the radial gradients of these properties.

\section{Conclusions}

In this pilot study we have obtained and analyzed IFU spectroscopy
of  six nearby spiral galaxies that hosted seven  SNe~Ia. 
For the data reduction we developed and tested 
a robust reduction pipeline. A set of tools that implement various methods to derive the properties
of the ionized gas and the stellar populations from the data-cube were also developed. This allowed us to generate 
 2D maps of the galaxies properties. The analysis of the maps showed that the quality of the data is  sufficient to
  accurately derive the properties of the ionized gas even in the outer low surface brightness parts of the galaxies.
 However, the parameters of the stellar populations are determined with much larger uncertainties. We showed that 
 analysis of azimutally averaged spectra at several de-projected galactocentric radii instead of the 2D maps provides a more robust way to 
 derive the radial dependencies of the stellar population  properties in galaxies with well-defined axial symmetry.
The main results of our study can be summarized as follows: 

\begin{itemize}

\item the six galaxies are quite massive with masses exceeding $2\times10^{10}\,M_{\sun}$;

\item the ionized gas and the stellar populations both indicate metallicities above the solar value;

\item five of the galaxies are currently forming stars at a rate of 1--5 $M_{\sun}$,yr$^{-1}$, which is typical for
spiral galaxies at $z\simeq0$. The sixth galaxy shows no signs of star formation;

\item the five star-forming galaxies have mean mass-weighted stellar age $\sim5$ Gyr and 
the passive one $\sim12$ Gyr;

\item four of the five star-forming galaxies show radial gradients of their ionized gas metallicity 
in the range from $-$0.022 to $-$0.058\,dex\,kpc$^{-1}$. These values are
 typical for other spiral galaxies in the local universe.  The fifth galaxy has 
a nearly uniformly distributed metallicity with a hint of a 
very low positive gradient of $+0.007$ dex\,kpc$^{-1}$;

\item the radial dependence of the stellar population properties can be more robustly derived if 
azimutally averaged spectra at several de-projected galactocentric radii are analyzed. By this analysis
 we found indication of low negative radial metallicity gradients of the 
stellar populations in some galaxies of up to $-$0.03\,dex\,kpc$^{-1}$. Given the large uncertainties with which the 
stellar metallicity is estimated, the significance of these gradient is difficult to assess;

\item in the five star-forming galaxies the fraction of young stellar populations increases until 4-8 kpc 
and shows signs of a subsequent decrease. In four of them the fraction of old stars monotonically 
decreases in the disk and one galaxy shows a more complex behavior.

\item the passive galaxy has mostly old stars, with a possible small fraction of younger stars in the 
central few kpc;

\item the kinematic analysis indicates that the galaxies are relaxed systems that most 
likely have not experienced recent a major merger;

\item most of the SNe in our sample are projected on regions with metallicity and star 
formation rates above the galaxy average, likely as a 
result of our target selection criteria and the  radial metallicity gradients; 

\item the BPT diagnostic diagram revealed that two of the galaxies host AGNs. Another galaxy is on the 
border between the AGNs and the  star-forming galaxies. Our analysis shows that the AGNs are not 
strong enough to affect the quantities derived from the total galaxy spectra.
This AGNs may not be recognized in studies of host galaxies of SNe~Ia at high redshift;

\item the correlation of the  SALT2 $x_1$ parameter of the SNe and the Hubble residuals with the various parameters
of the host galaxies did not lead to conclusive results. The low statistics and  the small ranges spanned by the 
galaxy parameters render such an attempt still premature.  We also note that the HRs of our SNe are  on average positive, although  their host galaxies have masses in the range where the other studies have shown negative HRs.

\end{itemize}

In conclusion, we have demonstrated the viability to study the host galaxies of SNe~Ia at low 
redshift using wide-field IFU spectroscopy at 4m-class telescopes. Intermediate-resolution spectra with sufficient S/N
can be obtained out to the outer low-surface brightness parts of the galaxies with a reasonably long exposure time of 
$\sim1.5$ hours.
Compared to a integrated spectroscopy and analysis of multi-color broad-band imaging, the IFU  spectroscopy provides
much more detailed information about the properties of the galaxies, e.g. metallicity and age gradients, detailed star formation histories, etc.
In principle, the S/N  of our data is sufficient to perform correlation analyses between the SN properties and the properties of the host galaxies at the location of the SN. 
However, our current sample it too small and suffers from strong selection biases to provide robust correlation results. 
The ongoing CALIFA survey may soon provide IFU spectroscopy of a larger sample of SNe~Ia host galaxies, which  will be 
a solid basis to further explore this path to study the SNe~Ia progenitors and improve SNe~Ia as distance indicators. 
We have tested the methodology and developed semi-automated tools that will allow expanding our analysis once the 
CALIFA data become available.

\begin{acknowledgements} 

V.S. acknowledges financial support from Funda\c{c}\~{a}o para a Ci\^{e}ncia
e a Tecnologia (FCT) under program Ci\^{e}ncia 2008. 
This work was partly funded by FCT with the research grant 
PTDC/CTE-AST/112582/2009 and a Ph.D. scholarship SFRH/BD/28082/2006, and under the 
Marie Curie Actions of the European Commission (FP7-COFUND).

This work has made use of the NASA/IPAC Extragalactic Database (NED),
NASA's Astrophysics Data System, and data products from SDSS and SDSS-II surveys.
Funding for the SDSS and SDSS-II has been provided by the Alfred P. Sloan Foundation, the Participating Institutions, the National Science Foundation, the U.S. Department of Energy, the National Aeronautics and Space Administration, the Japanese Monbukagakusho, the Max Planck Society, and the Higher Education Funding Council for England. The SDSS Web Site is \url{http://www.sdss.org/}.

\end{acknowledgements}


\Online

\begin{appendix}

\section{PMAS/PPAK instument details}
\label{ap:instr}

In this section we provide some details on the PPAK instrument that are relevant 
for the data reduction. For a full description of the instrument see \cite{2004AN....325..151V}, \cite{2005PASP..117..620R}, 
and \cite{2006PASP..118..129K}.

The PPAK fiber bundle consists of 382 
fibers with 2.7\arcsec diameter each, 331 of which (science fibers) are ordered
into a single hexagonal bundle that covers a field-of-view (FOV) of 
72\arcsec$\times$64\arcsec. Thirty-six additional fibers
form six mini-bundles of six fibers each (sky-bundles). The sky-bundles are evenly
distributed along a circle of radius $\sim90$\arcsec\ and face the edges of the
central hexagon  \citep[see Fig.5 in][]{2006PASP..118..129K}. The remaining 15 fibers are used for calibration and can only
be illuminated with the PMAS internal calibration unit.

The fibers are ordered into 12 slitlets. Each slitlet typically holds 28 science fibers, three
sky-fibers from three different sky-bundles and one calibration fiber. The three
sky-fibers are evenly distributed between the science fibers in the slitlet. The
12 slitlets are ordered to form a pseudo-slit. When projected onto the CCD in 2$\times$2 binned mode,
the separation between the spectra
along the cross-dispersion direction is $\sim$4.8 pixels and the full-width at
half-maximum (FWHM) of the spectral traces is $\sim2.5$ pixels. 
Between the slitlets there are gaps about two fibers wide, with the exception of the two central slitlets, which are
separated by wider gap of $\sim10$ fibers (see Fig.\,\ref{f:2d}; and also Figs.
13 and 14 in \cite{2006PASP..118..129K}). 

Because the fibers are circular,  only $\sim$65\% of the FOV is
spectroscopically sampled in a single exposure. Therefore, at least three suitably offset pointings are needed
to spectroscopically sample every point in the PPAK FOV.

\end{appendix}

\begin{appendix}

\section{Data reduction}

\label{ap:reduction}

The pre-reduction of the CCD images  was performed with IRAF and the remaining
reduction  with our own programs written in IDL. 

\subsection{Pre-reduction}

The PMAS 4k$\times$4k CCD is read  by four amplifiers and for each exposure four
separate FITS files are created. These were individually pre-reduced. The bias
frames showed no significant large-scale structures and the bias was
corrected by subtracting the
average value computed from the CCD overscans. The images were then trimmed, 
converted from ADUs into electrons using the gain values measured during the
commissioning of the CCD and  finally combined into a single 2D image, on which
the spectra were oriented roughly along the rows.

To create a master flat-field image, all halogen lamp exposures obtained
during the run where summed. The separation between the spectra
along the cross-dispersion direction is $\sim$4.8 pixels and the full-width at
half-maximum (FWHM) of the spectral traces is $\sim2.5$ pixels. As a result, the
intensity of the
pixels between the spectra is $\sim$20\% of the peak value \citep[e.g, see
Fig.18 in][]{2006PASP..118..129K}. Therefore, when all the images of the halogen
lamp spectra are combined, there will be enough counts even in between the
spectra. Moreover, because of the instrument flexure the positions of the
spectra change by up to $\sim1-2$ pixels, depending on the pointing of the
telescope.
As a result, in the combined image the intensities between the spectra were
$\sim80$\% of the peak values. Because the intensity changes smoothly along the
dispersion axis, each image row was
smoothed with running median in a window of 20 pixels 
to  normalize the master flat-field image. The flat-field image was
divided into its smoothed version to derive the final normalized flat-field
that contained only the pixel-to-pixel variations.

Because of the relatively long  exposures used, the images contain many 
pixels that are affected by cosmic ray (CR) hits. The complexity of PPAK images makes
it very difficult to use conventional methods for CRs rejection. We experimented
with different approaches and found that the CRREJ algorithm (developed for WFPC
on $HST$\footnote{We use the IDL implementation available in the IDL
Astronomy User's Library maintained by Wayne Landsman and available at
\url{http://idlastro.gsfc.nasa.gov/}}) gave the best results. We tried different
parameter settings, each time carefully examining the CR masks to verify that
no sky lines or galaxy emission lines were identified as CRs and
at
the same time as many as possible CRs were identified. It was occasionally necessary
to manually mark some CRs. Finally, the values of the CR hit pixels were
replaced with the values derived by interpolating the adjacent good pixels. We used
cubic-spline interpolation and the direction of the interpolation depended on
the extent of the  CR-affected region along the image rows and columns 

\subsection{Spectra tracing}

The continuum lamp exposures that were obtained before and/or after the
science exposures were used to trace the positions of the spectra on
the CCD. The spectra tracing was performed in two steps. First, 20 columns at the
middle of the image were
averaged and used to determine the location of all 382 spectra. Then moving left
and right and averaging 20 columns, the position of each spectrum at each CCD
column was determined using the peak positions of the previous
column as starting points. The positions of the maxima were determined by
fitting a parabola to
either the three or the four highest intensity points.  When the ratio between
the two
highest points was lower than 0.9, three points were used, in which case the
parabola passes exactly through all three points and the maximum can be
computed.  Otherwise, a least-squares fit was used to determine the maximum. 

The space  between the adjacent spectra is $\sim$4.8 pixels, which prevented us
from using a more accurate scheme to compute the positions of the spectra. The
traces determined at the first step were clearly oscillating with an
amplitude $\sim$1/3 pixels and it was necessary to additionally smooth them.
The traces were fit with a sixth-order
polynomial function and the results were stored in a FITS file.

Because of the mechanical flexures  of PMAS the positions of the spectra moved
slightly (at sub-pixel level) between the exposures, even within
the sequence of exposures of a given galaxy. This effect can be
accurately accounted for by a constant shift of the derived spectral traces. To
derive these shifts,  the average of 20 columns taken at the middle of the 2D
image was cross-correlated with the same average in the halogen 2D image that
was used to trace the spectra. In the subsequent reduction steps these shifts
were always accounted for. 

\subsection{Scattered light}

\begin{figure}[t]
\includegraphics [width=8.8cm]{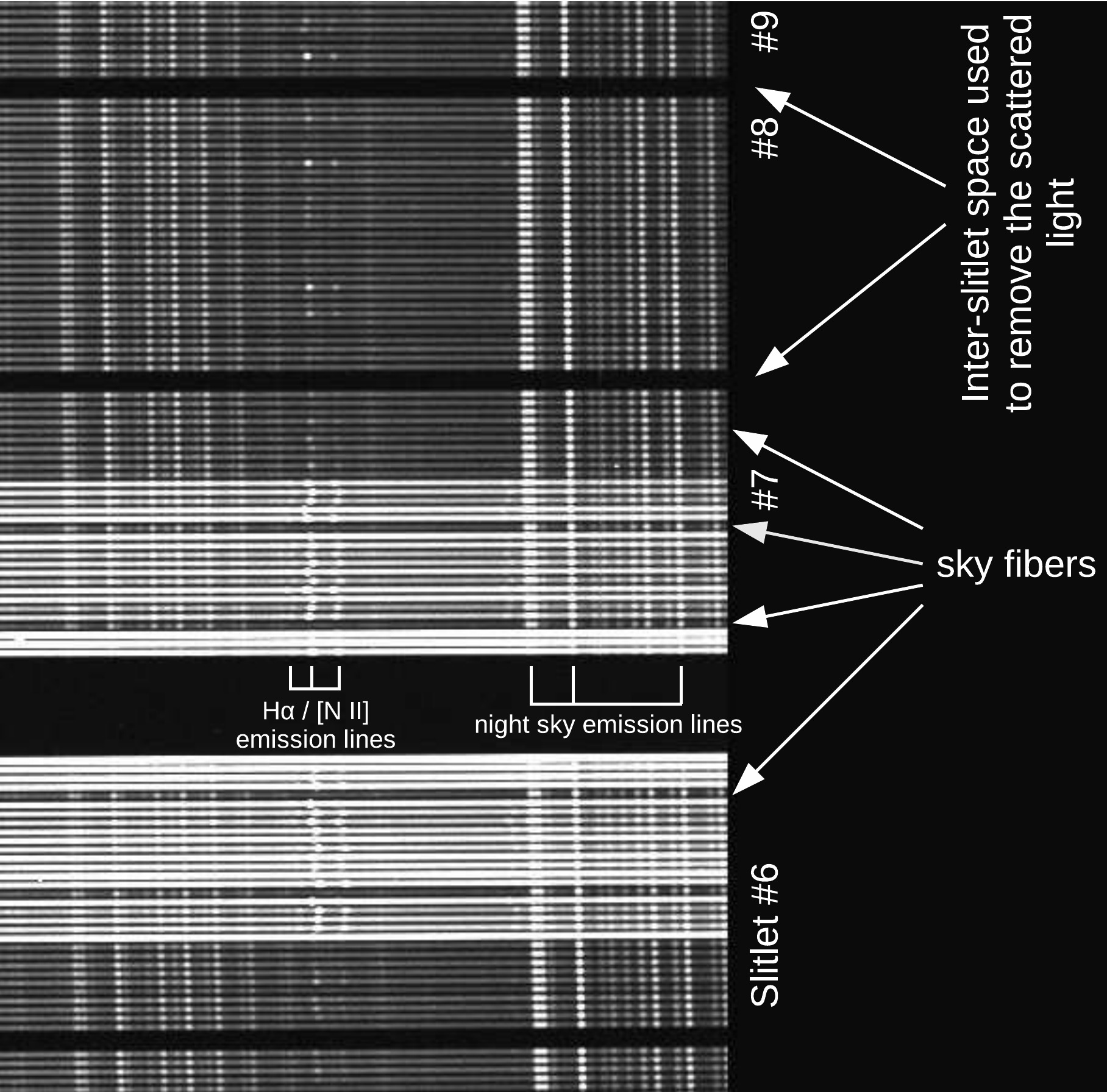}
\caption{Part of one of the science images that demonstrates the configuration
of the fibers on the CCD. The calibration fibers are not visible because they
are not illuminated during the science exposures. The calibration fibers are
normally last within the slitlets. Slitlet \#6 holds two calibration fibers, and two
more are placed at the beginning of the fiber sequence and one is at the end.}
\label{f:2d}
\end{figure}

A small amount of scattered light is present in the images and it needs to be
subtracted to achieve an accurate flux calibration. To derive a 2D model
of the scattered light, we used the gaps between the slitlets
(Fig.\,\ref{f:2d}), which contain no light from the object. 
Bands of five pixels width were extracted from the gaps (wider
band was extracted between the sixth and the seventh slitlets) using the
information from the spectra tracing. The bands were 
converted into 1D vectors by running median
in 5$\times$5 pixel window and the positions along the y-axis were also
computed. The scattered light pattern along the dispersion
axis is fairly complex and the attempts to directly fit a 2D polynomial surface
yielded no satisfactory results. Instead, the scattered light vectors 
were cubic-spline-interpolated along all image columns (in the spatial direction) 
to derive the amount of scattered light at each image pixel.  The resulting 2D model
image of the scattered light was smoothed by 50$\times$50 pixel running median
and subtracted from the original image.

\subsection{Spectra extraction and cross-talk correction}

The spectra were extracted from the 2D images using a simplified version of
the optimal extraction algorithm of \citet{1986PASP...98..609H}. In this
method the spatial profile of the spectral trace is determined from the data
themselves and is used to form a weighted sum of the pixels in a given aperture,
giving larger weights to the pixels that received
more light. In the conventional longslit spectroscopy the
profile is determined by the seeing and the telescope guiding, and may change
considerably from exposure to exposure. 
In the case of PPAK the spatial profiles are almost entirely determined by the
size of the fibers and 
have a stable shape that can be accurately described by a Gaussian. Thus, we 
determined the FWHMs of the Gaussians that describe the profiles of all 382
spectra and used this information to compute the weights for the optimal extraction.
Series of halogen lamp spectra were obtained and added 
 to increase the signal-to-noise ratio (S/N). For each image column 
382 Gaussians were simultaneously fitted to derive the FWHMs. The fitting was
performed 
with the MPFIT IDL package\footnote{available at
\url{http://cow.physics.wisc.edu/~craigm/idl/idl.html}} and special care was
taken to provide accurate starting values; otherwise the fit with so many
Gaussians would fail. The resulting 2D FWHM image (382 spectra by 2048 spectral
elements) was
somewhat noisy and was smoothed by running median within 11$\times$11 elements
window. The mean FWHM is 2.46 pixels and the covered range is 2.12-2.85.

With the positions of the spectra and the spatial profiles determined, the 
spectra were extracted with the optimal extraction algorithm. 
The separation between the spectra within the slitlets is fairly constant $\sim4.8-4.9$ pixels, 
both from spectrum to spectrum and along the trace. 
The extraction apertures were automatically computed as half the
average distance from closest spectrum. This means that for a given
spectrum the extraction aperture may be slightly asymmetric with respect to the
trace center and that different spectra may have
slightly different aperture sizes. 

Within a given slitlet the spectra are separated by only $\sim 2\times$FWHM, which leads to an
effect known as cross-talk, i.e. contamination by adjacent spectra\footnote{In the case
of PPAK the effect is small but we nevertheless chose to correct for it.}.
\citet{2006AN....327..850S} proposed the iterative \emph{Gaussian-suppression} technique
and we followed this approach with small modifications. After the first optimal
extraction,  the contamination from the adjacent spectra for a given spectrum was 
estimated as follows. Using the estimated flux, trace
positions and FWHMs of the Gaussians that describe the spatial profiles of the adjacent spectra, 
and taking into account the finite size of the extraction
apertures, we computed the height of the Gaussian functions that would have produced the
fluxes of the adjacent spectra were.  This was performed
 for all columns and the contamination of the adjacent spectra was subtracted.
Then a new optimal extraction was performed on the
corrected image using the same extraction aperture as before. This was performed for
all spectra and the process was iterated once more. 

\subsection{Wavelength calibration}

The HgNe and ThAr arc-lamp spectra were used to derive separate wavelength
calibrations for each individual spectrum. The arc-lamp spectra were extracted
in exactly the same way as the science ones and the positions of all Hg, Ne and Cd\footnote{few
Cadmium lines have also been identified in the spectra} lines with known
laboratory wavelengths were determined by a least-squares fit with Gaussian
function. A sixth-order polynomial fit was then used to derive accurate
pixel-to-wavelength transformations. The spectra extend from
about 3700\,\AA\ to 7000\,\AA. The bluest line in the HgNe spectrum is at
$\sim$4047\,\AA. Thus, the wavelength solution below $\sim$4000\,\AA\ may not be accurate. On
the other hard, the two reddest arc lines are at 6717\,\AA\ and 6929\,\AA. Because
of the field vignetting (see below) the 6929\,\AA\ line was not detected in all fibers
and as a result for those fibers the red part of the wavelength
solution may also be inaccurate. To work around this problem, the 15
calibration fibers that were illuminated with a ThAr lamp were used to improve the
wavelength solutions in the two extremes of the covered wavelength range. 
The dispersion over the wavelength range is between 1.5\,\AA\,pixel$^{-1}$ and
1.7\,\AA\,pixel$^{-1}$

\subsection{Background subtraction}

 The spectra extracted from fiber-fed or IFU spectrographs contain not only the
object light, but also the sky background, which needs to be subtracted. For 
faint objects the background may exceed the object flux by far and
accurate background subtraction is essential to obtain
reliable results. The PPAK design with the separate sky-fibers, 
which are distributed evenly between the science fibers, gives the flexibility
to apply
different approaches to the background subtraction. The following
method was found to work best. The sky-fibers were first examined to check if some of them were accidentally
contaminated by background objects. No such cases were found. Each element of
the 331 science spectra was labeled with a pair of numbers  ($\lambda_i$, $N$),
 its wavelength and fiber number as is recorded on the CCD, respectively. 
For a given element ($\lambda_i$, $N$) all sky-spectra were interpolated at a common wavelength
$\lambda_i$ using forth-order B-spline
interpolation\footnote{forth-order B-spline was used for all spectra
interpolations}, which has one of the best interpolation properties of all
known interpolation schemes \citep[e.g.,][]{bspline}. The interpolated
sky-spectra were fitted as a function of their fiber position with 
least-squares fifth order polynomial fit and the value of the sky at the element position $N$
was recorded. Repeating this for all elements gives a 2D model of the
sky-background, which was subtracted from the science fibers. The sky-subtracted 2D images 
were carefully examined and  no visible systematic
residuals were noticed, except at the position of the few strongest sky-lines. 

\subsection{Flux calibration}

\begin{figure}[t]
\includegraphics [width=8.8cm]{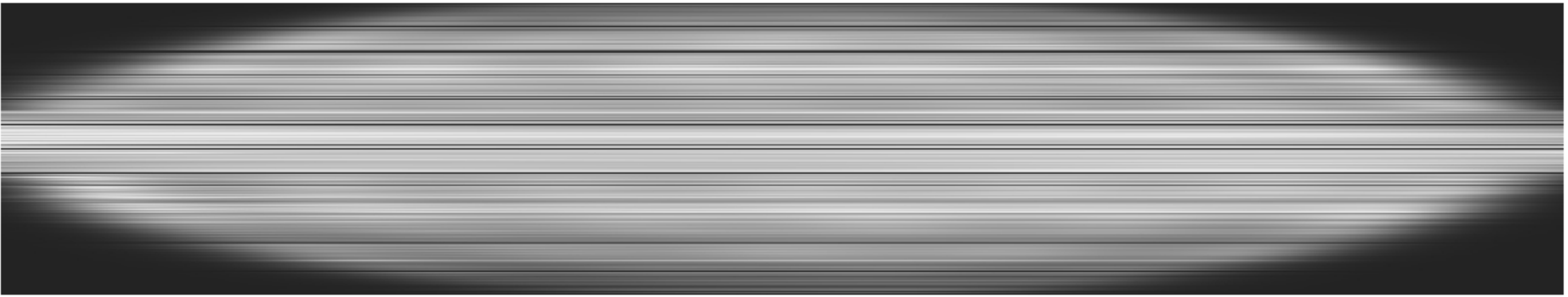}
\caption{Throughput map of the PPAK fibers. Note the vignetting in the corners.}
\label{f:vign}
\end{figure}

The flux calibration of the spectra was a two-step process: (i) normalization
for the fiber-to-fiber throughput variations and (ii) correcting for the wavelength-dependent
system response. To account for the fiber-to-fiber throughput variations the series of twilight sky 
exposures were used. The individual images were
added to increase the S/N and all spectra were extracted and wavelength-calibrated in the same way as
the science ones. Unfortunately, with the new larger 
CCD\footnote{the detector  was upgraded to 4k$\times$4k CCD in October 2009} the
images suffer from vignetting at the corners. The sensitivity of the parts of
the spectra that are projected at the CCD corners falls sharply as function
of the distance from the CCD center (Fig.\,\ref{f:vign}). The relation between
the on-sky fiber positions and the fiber position on
the CCD is such  that the most affected fibers are those adjacent to the central hexagon
of 37 fibers \citep[see Fig.\,7 in][]{2006PASP..118..129K} and not the outermost
fibers. 
The fibers of the central hexagon are projected at the middle of the CCD and are
practically un-affected by the vignetting. 

 To derive the relative throughput of the fibers, all spectra were interpolated on a
 common  wavelength frame. The 37 spectra of the central hexagon were
averaged and all spectra were divided by this average spectrum. The resulting ratios were
least-squares fitted with polynomials of different degrees (high orders were
needed for the
fibers most affected by the vignetting) to derive the throughput map of the fibers.
The map is shown in Fig.\,\ref{f:vign} and the vignetting at the corners is clearly visible. 
This map was used to equalize the
throughput of the fibers by dividing all science exposures by it.

During the analysis it was found that  for several of the fibers that were most severely 
affected by the vignetting the drop of the  throughput in the extreme blue and red parts 
could not be corrected well. This resulted in several spaxel spectra having incorrect relative 
flux calibration at the two extreme ends. This affected the spectrum fitting of these spaxels with {\tt STARLIGHT}. 
Because the H$\alpha$ flux was also affected, the estimation of the extinction was unreliable and hence part of the 
emission line analysis. The 'bad' spaxels were masked in the 2D maps of the affected properties.
It should be noted, however, that our primary gas metallicity indicator, the O3N2
method, was not affected because it only involves flux ratios of emission lines at close wavelengths.

\begin{figure*}[!th]
\sidecaption
\includegraphics [width=12cm]{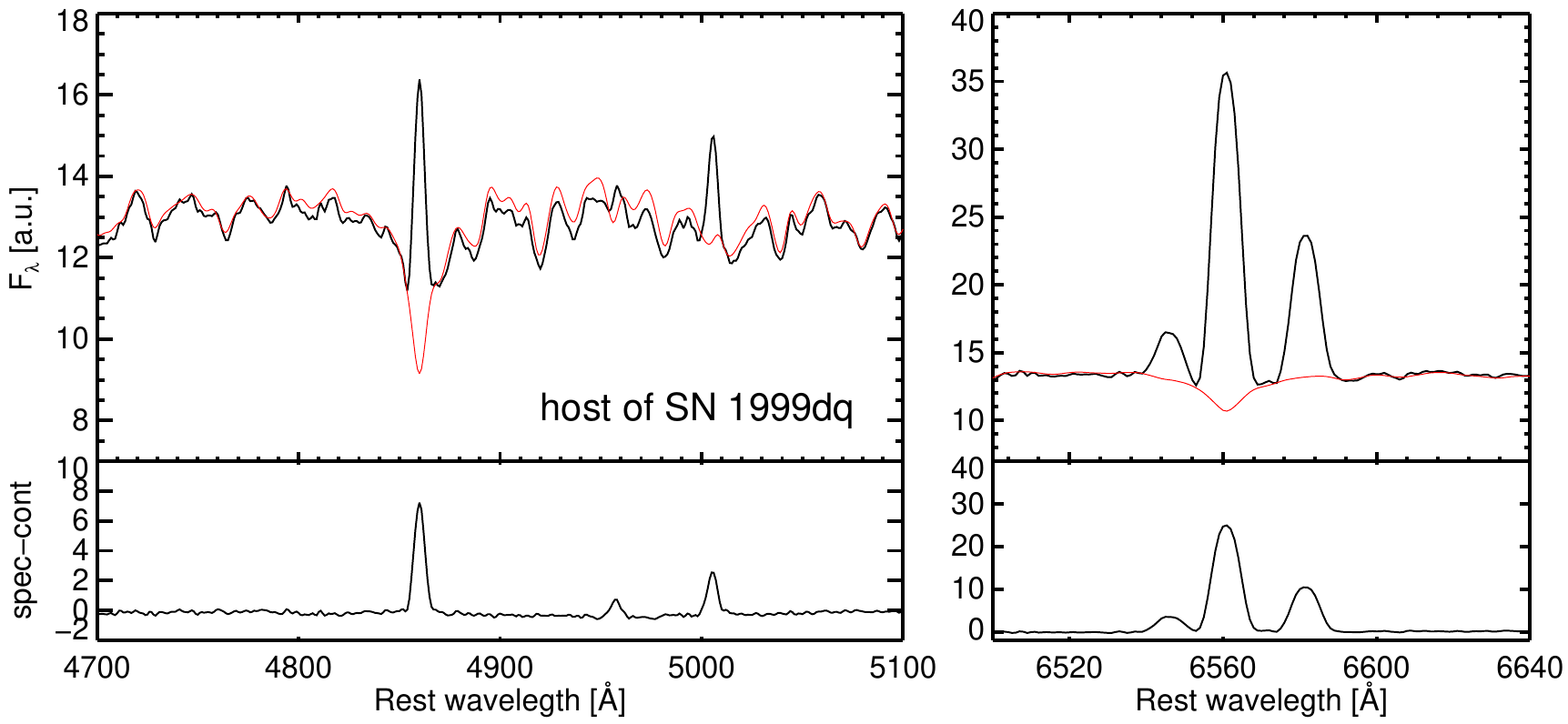} 
\caption{Example of the effect of the stellar spectrum subtraction in the regions of H$\beta$ and H$\alpha$ emission lines.
In the top panels we show the observed spectrum with the best fit overplotted. The bottom panels show the 
observed spectrum minus the best fit. Clearly, the fluxes of the emission line cannot be accurately measured without a proper subtraction of the stellar spectrum.}
\label{f:contsub}
\end{figure*}

 The two spectrophotometric standard stars that were observed on 15 November were used to
derive the system response. The spectra of the standards were extracted,
wavelength-calibrated, background-subtracted and fiber-throughput-corrected in the same way as
the rest of the science exposures. The presence of gaps between the fibers in
combination with the point source nature of the standards makes it difficult to
recover the whole flux of the star from a single exposure. In addition, because
of the differential atmospheric refraction (DAR) there may be significant  \emph{
differential} flux losses. By taking the
observations of the standards at five different points on the PPAK FOV, we aimed to minimize 
these effects. First, the
differential flux losses may cancel out when the five spectra are summed. 
Second, it gives more control, for example, too  discard a clearly
deviating observation.  

To minimize the flux losses instead of summing all
spectra that contain object flux, we adopted a different approach. The spectra were
interpolated at a common wavelength scale and each fiber was associated with its
 on-sky position. For each wavelength plane a symmetric 2D
Gaussian function (without constant term) was fitted and the flux was
taken to be the integral over the Gaussian. So extracted spectra were compared
with the result of simply suming the fibers. The spectra extracted with the
Gaussian fitting were clearly more consistent between the five pointings. With the
exception of one pointing for each standard, the systematic difference between
the spectra were less than $\sim3-4$\%. The deviating spectra for each star were
discarded and the remaining four were averaged. After correcting for
the atmospheric extinction using a mean extinction curve for Calar Alto
Observatory, the
ratio between the observed counts  and the tabulated fluxes of the standard were
fitted with a polynomial function to derive the sensitivity function of the
system. The sensitivities derived from the two standards agreed well and the average of the two 
sensitivities was used to calibrate all science
observations.

\subsection{Final data-cubes}

 The final step in the reduction was to combine the three pointings into a data
cube. The spectra were re-sampled into a common linear wavelength scale with
a sampling of 1.5\,\AA\ per element. Taking into account the offset between the
exposures, the circular fibers were mapped onto an output grid of square
2\arcsec$\times$2\arcsec pixels. The flux of a given fiber was distributed between
the output pixels according to the overlap area between the fiber and the output
pixels. No attempt was made to first shrink the fibers and emulate the Drizzle
algorithm used for $HST$ images \citep{2002PASP..114..144F}; with only three barely
overlapping pointing this would not gain any improvement. However, with five or
more carefully selected pointings it may be possible to use Drizzle to improve
the spatial sampling of the final cubes. 

Two important effects were taken into account two when creating the data-cubes -- the
non-photometric condition and the differential atmospheric refraction (DAR). To
account for the first, for each galaxy and pointing the total flux within an aperture of 
 12\arcsec\ centered on the galaxy core was computed and the observations of 
given galaxy were scaled to the exposure with the highest flux.  With one
exception all corrections were smaller than 15\%; the remaining pointing
required a correction factor of $\sim$2.

 The observations were obtained at an airmass of less that 1.3 and the effect of DAR was small, but
we nevertheless corrected for it. To compute the DAR we used 
the formulas from \citet{2005A&A...443..703S} and the ambient conditions
as recorded in the FITS image headers. To combine the three
pointings, we shifted  the positions of the fibers for each wavelength slice
along the $x$ and $y$ axes by the values computed from the value of the DAR at
that wavelength (with respect to the refraction at 6300\,\AA) and the effective
parallactic angle of the observations \citep[see also][]{1990ESOC...34...95W}.
For both the DAR-corrected and un-corrected data-cubes, the positions of the galaxy nuclei 
were computed in the blue (average in 3800-4000\AA\AA) and red (6600-6800\AA\AA)
wavelength ranges. The two positions measured in the DAR-corrected cubes matched very well, 
but not in the un-corrected cubes, which shows that the DAR
was correctly accounted for.

 Three of the galaxies in our sample also have SDSS spectroscopy. This allowed us 
to check the \emph{relative} flux calibration of our spectra.
To emulate the SDSS spectroscopy, we extracted from our data-cubes 
the flux within an aperture of 3\arcsec\ diameter centered on the galaxy nucleus and
scaled it to match the mean flux level of the SDSS spectra. 
 The comparison between the SDSS spectra and those extracted from our date-cubes is shown in Fig.\,\ref{f:sdss}. 
It shows that the \emph{relative}
 flux calibration of our spectra is excellent and matches SDSS to within a few percent.

To set the absolute flux scale of the data-cubes we used the SDSS imaging.
SDSS $g$ and $r$ magnitudes of the galaxies were 
 computed within an aperture of 20\arcsec\ diameter. Spectra within the same aperture size 
 were extracted from the data-cubes and synthetic $g$ and $r$ magnitudes were computed. 
 The $g$ and $r$ scale factors that made the synthetic magnitudes match the observed 
 ones were computed and the average of the two was applied to the data-cubes. 
 It should be noted that the  $g$ and $r$ scale factors coincided to within a few percent,
 which supports our conclusion that the \emph{relative} flux calibration is accurate.

\end{appendix}

\begin{appendix}

\section{Measurement of emission line fluxes}
\label{ap:lines}

Five of the six galaxies show strong
nebular emission lines. This enabled us to study the properties of the
gas phase in the galaxies through measuring the fluxes of the most
prominent emission lines
[\ion{O}{ii}]\,$\lambda$3727, H$\beta$,
[\ion{O}{iii}]\,$\lambda\lambda$4959/5007,  H$\alpha$, and
[\ion{N}{ii}]\,$\lambda\lambda$6549/6584, and when possible also
 [\ion{S}{ii}]\,$\lambda\lambda$6716/6731. 
In spectra of galaxies the
nebular emission lines are superimposed on the underlying stellar
continuum. The stellar absorption lines can bias the measurement of the
emission line fluxes, an effect that is especially prominent in H$\beta$ (Fig.~\ref{f:contsub}). Therefore,
the stellar continuum needs to be subtracted first
to measure the emission line fluxes  accurately. For this we used the STARLIGHT software
\citep{2005MNRAS.358..363C}. Briefly, after masking the regions of known
nebular emission lines, telluric absorptions, and strong night-sky emission
lines, the spectrum was fitted with a linear combination of model spectra
of single stellar populations (SSP) of different ages and metallicities. The best
fit was subtracted to derive the pure emission line spectrum, where the emission
line fluxes can be measured without bias, see Fig.~\ref{f:contsub}. The parameters of the fitted SSPs can be
used to derive the properties of the galaxy stellar population, but here we regarded the fits as a means to
subtract the underlying stellar absorption. 
All spectra that had S/N higher than 5 at
$\sim4600$\AA\ were fitted and the emission line fluxes were measured on the
continuum-subtracted spectrum. For the remaining spectra the measurements were
made without continuum subtraction. A weighted non-linear least-squares fit with a
single Gaussian plus a linear term was performed for each emission line, and the
area below the Gaussian was taken as an estimate of the flux. The weights were
the error spectrum produced during the spectral extraction. Including the linear term 
is necessary to account for possible  systematic residuals due to imperfect 
subtraction of the  the stellar absorption spectrum.
The close lines
H$\alpha$, [\ion{N}{ii}]\,$\lambda$6549 and [\ion{N}{ii}]\,$\lambda$6584 were
simultaneously fitted.
[\ion{O}{ii}]\,$\lambda$3727 is a
blend of two lines -- [\ion{O}{ii}]\,$\lambda\lambda$3726.04/3728.80. This
blend was fitted with two Gaussians with equal width and a fixed separation of
 2.76\,\AA. In addition, a robust standard deviation of the adjacent continuum
was also measured from the regions more then 3$\sigma$ away from the line centers.

The uncertainty of the line fluxes was estimated by propagating the
uncertainties of the fitted amplitude and $\sigma$ of the Gaussians. To check the
reliability of this estimate, we performed Monte Carlo simulations. We assumed
that the lines have a Gaussian shape with $\sigma=2$, representative for
our spectra, and that they were superimposed on constant background
with standard deviation one and mean zero. The noise across the line was assumed to
follow a Poisson distribution -- if a given datum has $N$ counts, the
1$\sigma$ uncertainty of this datum is $(1+N)^{1/2}$. This noise model
was used to generate pseudo-random numbers, which were added to the line and the
line was fitted (weighted by the noise model). This was
repeated 1000 times and
the mean values and the standard deviation of $\sigma$, the
amplitude, and the area below the Gaussian were computed. This procedure was repeated for
Gaussians with 
amplitudes 2 to 1000, i.e. exceeding the continuum noise by the same numbers.

The results of these simulations show that propagating the errors of 
the fitted parameters to the total flux overestimates the uncertainty by
about a factor of 2. The simulations also showed that the area
below the Gaussian can be recovered with $\sim30$\% accuracy and without
significant bias for lines that exceed the continuum noise by only three times. 
The S/N of the measured line flux was tabulated as a function of the ratio of 
the \emph{fitted} amplitude of the Gaussian to 
the standard deviation of the adjacent continuum and was used to
assign realistic errors to the line fluxes measured by fitting a Gaussian
function. 
For example, a 10\% accuracy is achieved when the amplitude of the Gaussian is 20 times the
continuum noise. 
A possible concern is that the shape of the lines in the real spectra
may deviate from Gaussian, in which case additional uncertainty will be introduced. Close inspection
of the line fits revealed that the line shape in our spectra is well represented by a Gaussian function and
we do not expect problems related to a non-Gaussian line shape.

\end{appendix}

\begin{appendix}
\section{Online figures}

\end{appendix}
\end{document}